\documentclass[acmsmall]{acmart}
\AtBeginDocument{%
  }

\setcopyright{rightsretained}
\acmJournal{TOSEM}
\acmYear{2025} \acmVolume{1} \acmNumber{1} \acmArticle{1} \acmMonth{1} \acmPrice{}\acmDOI{10.1145/3748505}


\usepackage{amsmath,amsfonts}
\usepackage{algorithmic}
\usepackage{graphicx}
\usepackage{textcomp}
\usepackage[ruled,vlined,linesnumbered]{algorithm2e}
\usepackage{multirow}
\usepackage{adjustbox}
\usepackage{pifont}
\usepackage{threeparttable}
\usepackage{booktabs}
\usepackage{bbding}
\usepackage{tablefootnote}
\usepackage{hyperref}
\usepackage{tcolorbox}
\usepackage{enumitem}
\usepackage{enumerate}
\usepackage{subfig}
\usepackage{url}
\usepackage{soul}
\usepackage{balance}
\usepackage{makecell}
\hypersetup{
colorlinks=true,
urlcolor=black,
linkcolor=black,
citecolor=black
}

\usepackage{ulem}
\usepackage{collcell}

\newcommand{\del}[1]{}

\newcolumntype{S}{>{\collectcell\del}c<{\endcollectcell}}

\begin{document}

\definecolor{imp}{RGB}{153,236,235}

\newcommand{\tool}{TELPA}
\newcommand{\tech}{TELPA}

\newcommand{\coda}{CODAMOSA}
\newcommand{\sbst}{Pynguin}
\newcommand{\evo}{EvoSuite}
\newcommand{\llm}{CHATTESTER}
\newcommand{\tester}{CHATTESTER}

\newcommand{\llmrevised}{CHATTESTER$^{\text{phi}}$}
\newcommand{\codarevised}{CODAMOSA$^{\text{phi}}$}

\newcommand{\llmsbst}{CHATTESTER$_p$}
\newcommand{\llmcoda}{CHATTESTER$_c$}

\newcommand{\llmsbstrevised}{CHATTESTER$_{p}^{\text{phi}}$}
\newcommand{\llmcodarevised}{CHATTESTER$_{c}^{\text{phi}}$}

\newcommand{\techzero}{\tech$_{\textit{scratch}}$}

\newcommand{\deepseekbase}{\tool$_{\textit{ds}}$}

\newcommand{\deepseeksbst}{\tool{}$_{\textit{ds}}^p$}
\newcommand{\deepseekcoda}{\tool{}$_{\textit{ds}}^c$}

\newcommand{\gptsbst}{\tool{}$_{\textit{gpt}}^p$}

\newcommand{\pynguinbase}{\tool$_{p}$}
\newcommand{\codamosabase}{\tool$_{c}$}
\newcommand{\evosuitebase}{\tool$_{e}$}
\newcommand{\llmbase}{\tool$_{ct}$}

\newcommand{\pynguinstall}{\sbst{}$_{\textit{bn}}$}
\newcommand{\codastall}{\coda{}$_{\textit{bn}}$}
\newcommand{\codastallrevised}{\coda{}$_{\textit{bn}}^{\text{phi}}$}

\newcommand{\nocall}{\tool$_{\textit{nba}}$}
\newcommand{\nobranch}{\tool$_{\textit{nfa}}$}
\newcommand{\noshot}{\tool$_{\textit{nce}}$}
\newcommand{\nofeedback}{\tool$_{\textit{nf}}$}
\newcommand{\nofiltering}{\tool$_{\textit{npf}}$}
\newcommand{\nocot}{\tool$_{\textit{ncot}}$}
\newcommand{\randomshot}{\tool$_{\textit{rce}}$}

\newcommand{\nocallsbst}{\tool$_{\textit{nba}}^p$}
\newcommand{\nobranchsbst}{\tool$_{\textit{nfa}}^p$}
\newcommand{\noshotsbst}{\tool$_{\textit{nce}}^p$}
\newcommand{\nofeedbacksbst}{\tool$_{\textit{nf}}^p$}
\newcommand{\nofilteringsbst}{\tool$_{\textit{npf}}^p$}
\newcommand{\nocotsbst}{\tool$_{\textit{ncot}}^p$}
\newcommand{\randomshotsbst}{\tool$_{\textit{rce}}^p$}

\newcommand{\nocallcoda}{\tool$_{\textit{nba}}^c$}
\newcommand{\nobranchcoda}{\tool$_{\textit{nfa}}^c$}
\newcommand{\noshotcoda}{\tool$_{\textit{nce}}^c$}
\newcommand{\nofeedbackcoda}{\tool$_{\textit{nf}}^c$}
\newcommand{\nofilteringcoda}{\tool$_{\textit{npf}}^c$}
\newcommand{\nocotcoda}{\tool$_{\textit{ncot}}^c$}
\newcommand{\randomshotcoda}{\tool$_{\textit{rce}}^c$}

\newcommand{\homepage}{https://anonymous.4open.science/r/TELPA-CF78}

\newcommand{\jj}[1]{{\color{orange}[Junjie: #1]}} 
\newcommand{\bin}[1]{{\color{blue}[Bin: #1]}} 
\newcommand\yc[1]{\textcolor{red!50}{[yc: #1]}}

\newcommand{\colorblue}{\color{black}} %
\newcommand{\ins}[1]{{\colorblue{#1}}}

\newcommand{\delminor}[1]{} 
\newcommand{\colorblueminor}{\color{black}} %
\newcommand{\insminor}[1]{{\colorblueminor{#1}}}

\title{Advancing Code Coverage: Incorporating Program Analysis with Large Language Models}

\author{Chen Yang}
\orcid{0000-0003-0759-940X}
\affiliation{%
\department{College of Intelligence and Computing}
  \institution{Tianjin University}
  \city{Tianjin}
  \country{China}}
\email{yangchenyc@tju.edu.cn}

\author{Junjie Chen}
\orcid{0000-0003-3056-9962}
\authornote{Corresponding author (junjiechen@tju.edu.cn)}
\affiliation{%
\department{College of Intelligence and Computing}
  \institution{Tianjin University}
  \city{Tianjin}
  \country{China}}
\email{junjiechen@tju.edu.cn}

\author{Bin Lin}
\affiliation{%
  \institution{Hangzhou Dianzi University}
  \city{Hangzhou}
  \country{China}}
\email{b.lin@hdu.edu.cn}

\author{Ziqi Wang}
\affiliation{%
\department{College of Intelligence and Computing}
  \institution{Tianjin University}
  \city{Tianjin}
  \country{China}}
\email{wangziqi123@tju.edu.cn}

\author{Jianyi Zhou}
\affiliation{%
  \institution{Huawei Cloud Computing Technologies Co., Ltd.}
  \city{Beijing}
  \country{China}}
\email{zhoujianyi2@huawei.com}

\begin{abstract}
Automatic test generation plays a critical role in software quality assurance. While the recent advances in Search-Based Software Testing (SBST) and Large Language Models (LLMs) have shown promise in generating useful tests, these techniques still struggle to cover certain branches. Reaching these hard-to-cover branches usually requires constructing complex objects and resolving intricate inter-procedural dependencies in branch conditions, which poses significant challenges for existing techniques. 
In this work, we propose \tool{}, a novel technique aimed at addressing these challenges. 
Its key insight lies in extracting real usage scenarios of the target method under test to learn how to construct complex objects and extracting methods entailing inter-procedural dependencies with hard-to-cover branches to learn the semantics of branch constraints. 
To enhance efficiency and effectiveness, \tool{} identifies a set of ineffective tests as counter-examples for LLMs and employs a feedback-based process to iteratively refine these counter-examples.
Then, \tool{} integrates program analysis results and counter-examples into the prompt, guiding LLMs to gain deeper understandings of the semantics of the target method and generate diverse tests that can reach the hard-to-cover branches. 
Our experimental results on 27 open-source Python projects demonstrate that \tool{} significantly outperforms the state-of-the-art SBST and LLM-enhanced techniques, achieving an average improvement of 34.10\% and 25.93\% in terms of branch coverage.

\end{abstract}


\maketitle

\section{Introduction}
\label{sec:intro}
Automatic test generation holds significant importance in software quality assurance, which aims to efficiently detect software bugs by automatically covering various behaviors of the software under test.
Recently, many test generation techniques have been proposed, among which Search-Based Software Testing (SBST)~\cite{pynguin,randoop,evosuite} is one of the most widely-studied categories due to its effectiveness.
Typically, SBST techniques leverage heuristic search algorithms to explore the vast test space by mutating already-generated tests with the guidance of maximizing code coverage.
While receiving extensive attention, they are still unable to reach specific parts of code. %
For instance, they often fail to generate tests to cover the branches that can be triggered only by some scenario-specific values \del{(e.g., specific file names, version strings) due to the lack of the ability of understanding the code semantics~\cite{codamosa}}\ins{that may require deep understanding of the code semantics}.

The recent advancement of Large Language Models (LLMs) provides new opportunities for tackling these problems, and several LLM-based techniques have been proposed to generate effective test cases~\cite{codamosa,chattester}.
These techniques utilize the code comprehension ability of LLMs by incorporating the source code of the target method to be tested and some contextual information (such as other methods around the target method) into the prompt.
While the limitation of SBST can be relieved to some degree with the help of LLMs, the automatically generated tests still tend to be generic and only able to cover the branches without complicated constraints~\cite{codeaware}.
That is, many\del{ hard-to-cover} branches are still left for manual test design based on experts' domain knowledge, leading to significant costs.
An automated approach that can further improve the coverage of these\del{ hard-to-cover} branches can largely reduce the testing effort.  %

In practice, two major challenges stand out when creating such approaches.
First, some branch constraints involve objects with complex construction processes. 
Specifically, the construction of some objects relies on other (complex) types of objects, ultimately necessitating a specific sequence of constructor invocations. 
\ins{
For example, the branch condition at Line 5 in Figure~\ref{fig:motivation_1_target} requires a \texttt{field} object with a specific type (e.g., \texttt{Array}) and a specific attribute value (e.g., \texttt{items} should be a \texttt{(tuple, list)} with the type \texttt{typing.Union[Field, typing.Sequence[Field]]}). 
Constructing such a \texttt{field} object is challenging because it requires understanding the dependencies between multiple objects: (1) constructing valid \texttt{items} (which must be either a single \texttt{Field} object or a sequence of \texttt{Field} objects), and (2) creating a valid \texttt{field} object, ensuring that the \texttt{items} attribute has the correct type. This process involves multiple steps and dependencies that are difficult for existing techniques to handle without a deep understanding of the code semantics.
}
As indicated by a previous study~\cite{chattester}, the tests generated by LLMs exhibit a relatively low compilation success rate (about 39\%), \del{highlighting a deficiency in the ability of LLMs in this regard.}\ins{which directly impacts their effectiveness in achieving code coverage. This highlights a deficiency in the ability of LLMs to generate effective tests in such scenarios.}
Indeed, state-of-the-art LLM-based techniques (e.g., \coda{}~\cite{codamosa}) fail to produce valid objects for such scenarios, let alone the objects with specific attribute values required by branch constraints.
Second, some branch constraints entail intricate inter-procedural dependencies. 
That is, the outcome of certain conditions within the constraints are determined by the execution results of a series of method invocations. 
\ins{
For instance, the branch condition at Line 3 in Figure~\ref{fig:motivation_2_target} depends on the result of the method \texttt{is\_magic}. This method is defined elsewhere in the project and may invoke other methods, creating a chain of inter-procedural dependencies. To cover this branch, the test must correctly simulate the action that \texttt{is\_magic} performs and provide the test data that satisfy the condition. This requires a deep understanding of the code semantics to ensure that the generated test accurately reflects the dependencies and produces the correct result for the branch condition.
}
In such scenarios, merely providing source code of the target method or some coarse-grained contextual information (the common practice of existing LLM-based techniques) is insufficient to enable LLMs to comprehend the semantics conveyed by a sequence of invoked methods, thereby struggling to generate effective tests.
\ins{
Therefore, we define ``hard-to-cover branches'' as those that meet either of the following criteria: (1) branches requiring values derived from complex object construction. For example, if a branch condition checks whether an object's attribute matches specific conditions (such as type, format, and other non-constant conditions), the test must first construct the object with the appropriate attribute to cover the branch; or (2) branches with complex inter-procedural dependencies. For example, if a branch condition depends on the result of a method invocation, which itself relies on other method invocations, the test must accurately simulate the sequence of method calls and their interactions to cover the branch.
}

To tackle these challenges\ins{ in covering \delminor{hrad-to-cover}\insminor{hard-to-cover} branches}, we propose a novel LLM-based test generation technique (called \textbf{\tech{}} -- \textbf{TE}st generation via \textbf{L}lm and \textbf{P}rogram \textbf{A}nalysis), which focuses on reaching hard-to-cover branches via program-analysis-enhanced prompting.
Given that complex objects in branch constraints are often passed to the target method as parameters, \tech{} addresses the first challenge by gathering invocations of the target method within the software module. 
These existing invocations might contain the whole process of constructing these objects.
Specifically, \tech{} conducts \del{backward method-invocation analysis}\ins{\textbf{object construction analysis}},\ins{ which traces the sequence of method calls
leading to the target method, focusing on understanding how objects are constructed
and passed to the target method. This analysis} \del{starting}\ins{starts} from each invocation of the target method, to extract all sequences of method invocations with the target method as the endpoint. 
Then, \tech{} prompts LLMs to generate tests by using different paths to enter the target method.
By exploring different entry paths, \tech{} not only facilitates the construction of valid objects but also captures various usage scenarios 
in practice (that may produce specific attribute values required by branch constraints), thus enhancing test diversity and improving coverage.

To address the second challenge arising from complicated inter-procedural dependencies in branch constraints, \tech{} performs \del{forward method-invocation analysis}\ins{\textbf{branch dependency analysis}}\ins{, which determines the methods invoked within the branch conditions of the target method, focusing on understanding the inter-procedural dependencies that influence branch outcomes}.
This analysis begins from each variable and invoked method in the constraints for a target branch, aiming to extract all associated methods.
By incorporating the source code of these methods in their invocation order as the prompt, \tech{} can elicit LLMs to gain a deeper understanding of the semantics pertaining to the target branch, thereby generating effective tests. This provides precise contextual information concerning the target branch and it is more effective than directly supplying the source code of all methods within the software module to LLMs, as the latter could introduce excessive noise irrelevant to the target branch and confuse LLMs during test generation. 

Due to the non-negligible cost of utilizing LLMs, we define the usage scenario of \tech{} as its activation when the existing (lightweight) test generation tool (e.g., \sbst{}~\cite{pynguin}) fails to increase test coverage within a certain timeframe. 
In other words, \tech{} is only employed for hard-to-cover branches to ensure cost-effectiveness.
To enhance efficiency further, \tech{} samples a diverse set of already-generated tests as counter-examples. 
By incorporating these counter-examples into the prompt, \tech{} can instruct LLMs to generate tests that diverge from them, as these counter-examples have been recognized as ineffective for these hard-to-cover branches.
In particular, the overall test generation process with \tech{} is structured as a feedback-based process. 
This design allows for updating counter-examples based on the most recent coverage results, thereby enhancing test generation effectiveness.

We conducted an extensive evaluation on 27 open-source Python projects, which have been widely used in previous studies~\cite{codamosa,pynguin}.
Our results show that \tech{} significantly outperforms both the state-of-the-art SBST \del{technique}\ins{tool} (i.e., \sbst{}~\cite{pynguin}) and the state-of-the-art LLM-based\ins{/LLM-enhanced} techniques (i.e., \coda{}~\cite{codamosa} and \llm{}~\cite{chattester}).
On average, \tech{} achieves 34.10\%, 25.93\%, and 21.10\% higher branch coverage than them respectively, given the same testing time budget.
Our ablation study also confirms the contribution of each main component in \tech{}, including
\del{backward method-invocation analysis}\ins{object construction analysis} for relieving the challenge arising from complex object construction, 
\del{forward method-invocation analysis}\ins{branch dependency analysis} for relieving the challenge arising from intricate inter-procedural dependencies,
counter-examples sampling and coverage-based feedback for improving the overall testing efficiency.
Our investigation into various configurations of \tech{} confirms that our designed usage scenario for \tech{} strikes a balance between effectiveness and efficiency and \tech{} consistently achieves higher coverage than other settings. %

The main contributions of our study are as follows. %
\begin{itemize}
    \item We introduce \tech{}, a novel LLM-based test generation technique designed to enhance the coverage of hard-to-cover branches through program-analysis-enhanced prompting.

    \item \tech{} can \del{well }address the challenges arising from complex object construction and intricate inter-procedural dependencies \ins{well }by incorporating both \del{backward and forward method-invocation analysis}\ins{object construction and branch dependency analysis} within \tech{}.

    \item The feedback-based test generation process in \tech{} improves the efficiency and effectiveness of test generation by guiding LLMs with diverse counter-example tests.
    
    \item We present an extensive evaluation of \tech{} across 27 open-source projects, demonstrating its effectiveness in comparison to \del{a}\ins{the} state-of-the-art SBST \del{technique}\ins{tool} and two state-of-the-art LLM-based\ins{/LLM-enhanced} techniques.
\end{itemize}

\section{Motivating Example}
\label{sec:motivation}

In this section, \delminor{we demonstrate the limitations of the state-of-the-art}\insminor{we demonstrate the limitations of state-of-the-art automated test generation tools for Python, namely} \sbst{}~\cite{pynguin}, \coda{}~\cite{codamosa}, and \tester{}~\cite{chattester} with two real examples to motivate our study.

\begin{figure}[t]
  \centering
  \includegraphics[width=0.68\linewidth]{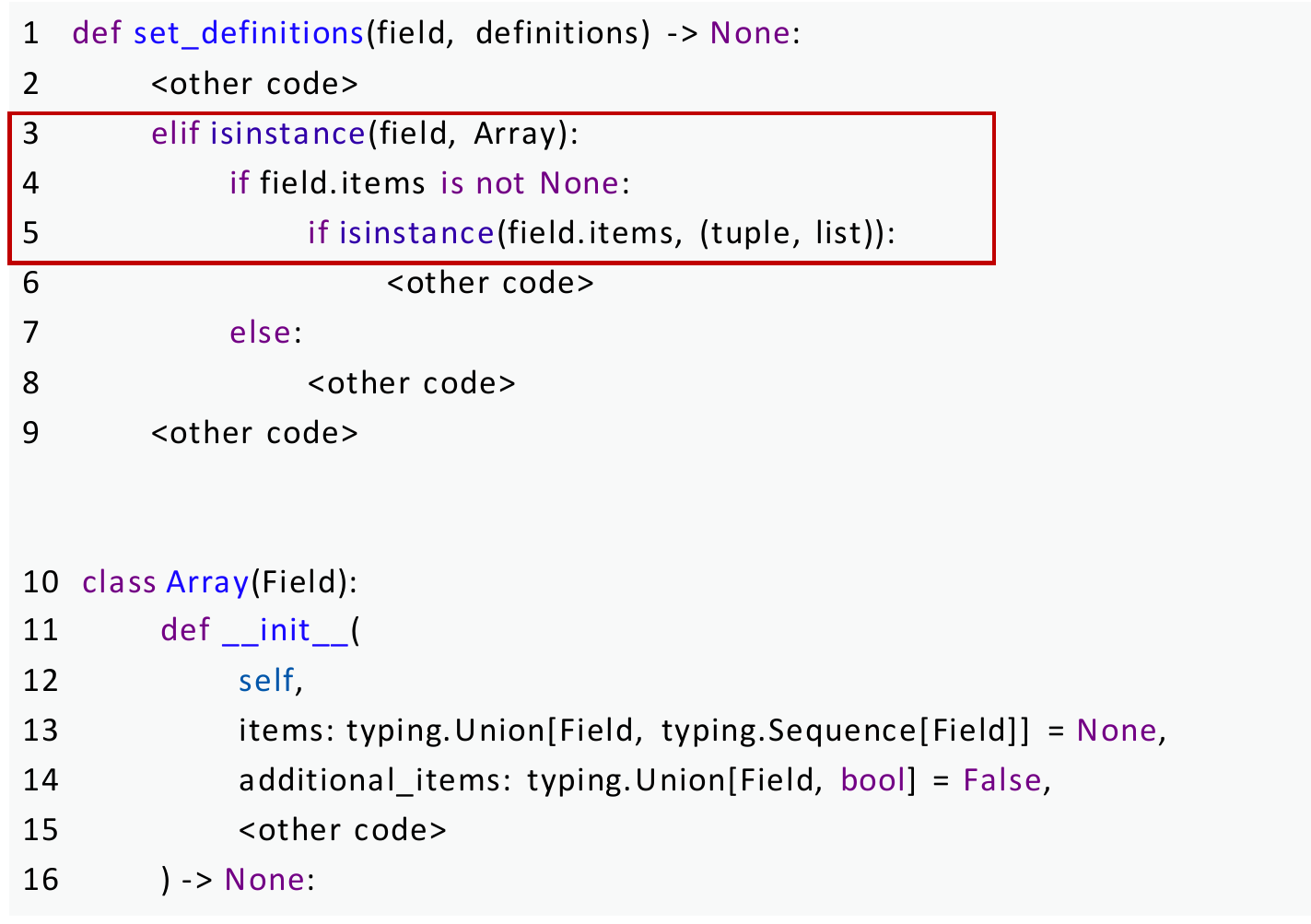}
  \caption{Target method \textit{set\_definitions}}
  \label{fig:motivation_1_target}
\end{figure}

Figure~\ref{fig:motivation_1_target} shows the simplified code snippet of the target method \texttt{set\_definition} from the \texttt{typesystem} 
project\footnote{https://github.com/encode/typesystem}. With a time budget of 20 minutes, \sbst{}, \coda{}, and \tester{} all fail to reach line 6, namely the condition for the {\tt if} statement in line 5 never returns true. In fact, to make the condition true, the generated test needs to first construct an \texttt{Array} object (line 3), and then assign a proper value to its \texttt{items} attribute. More specifically, \texttt{items} should be either a tuple or a list of \texttt{Field} objects (line 4), according to the type hints from the constructor of \texttt{Array} (line 13).  %
Constructing such a sophisticated \texttt{Array} object is certainly challenging for automatic test generation techniques, \sbst{}, \coda{}, and \tester{} are no exception. 

\begin{figure}[t]
  \centering
  \includegraphics[width=0.68\linewidth]{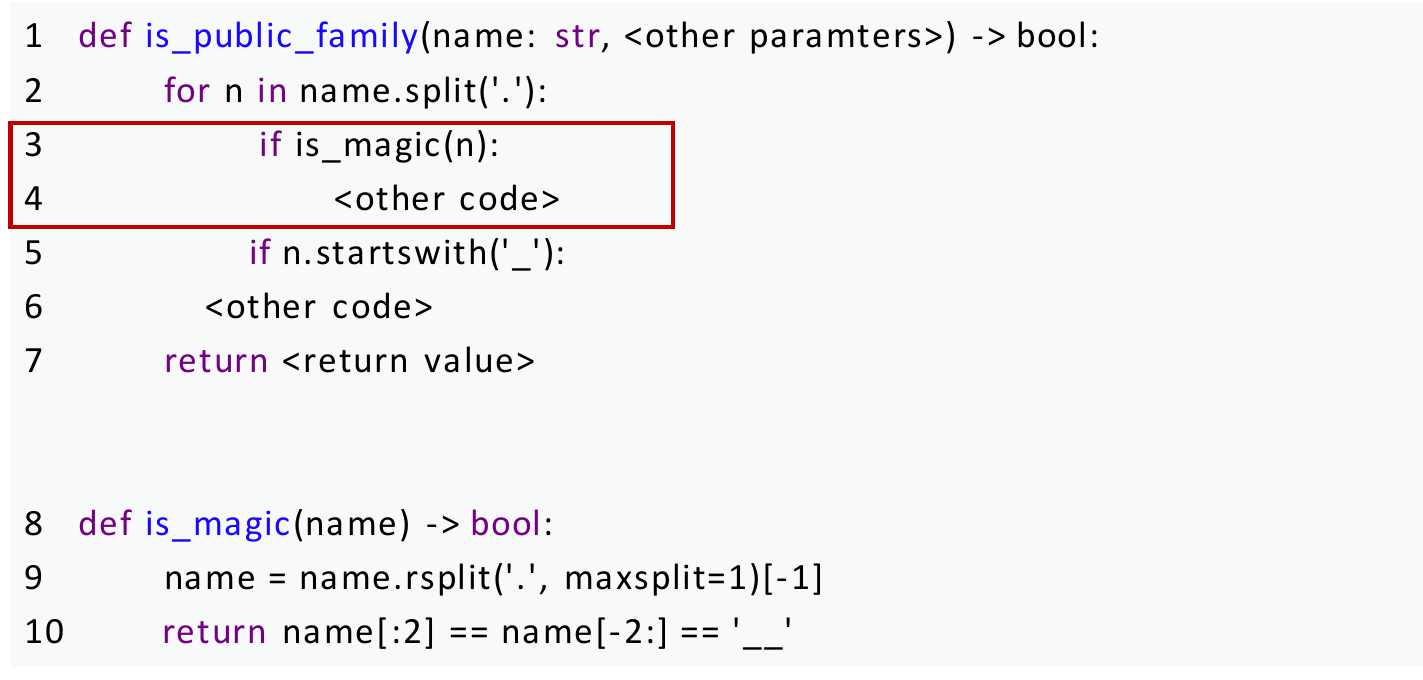}
  \caption{Target method \textit{\_is\_public\_family}}
  \label{fig:motivation_2_target}
\end{figure}

Another type of hard-to-cover branches are caused by complicated inter-procedural dependencies, as illustrated in Figure~\ref{fig:motivation_2_target}. In this case, our target method is \texttt{is\_public\_family} from the \texttt{apimd} project\footnote{https://github.com/KmolYuan/apimd}. \sbst{}, \coda{}, and \tester{} are unable to guide the condition in line 3 to return true. 
The challenge here is that the outcome of the {\tt if} condition depends on the method \texttt{is\_magic}, which is defined elsewhere in the project
and might further invoke other methods. 
Those complicated inter-procedural dependencies are often hard to understand and analyze for test generation techniques.

\ins{According to our analysis, our evaluation benchmark, which consists of 486 modules from 27 Python projects, contains 6,559 branches, of which 3,979 (60.7\%) are classified as hard-to-cover. Similar challenges are also prevalent in Java, as evidenced by the Defects4J benchmark, which contains 9,391 branches, of which 5,494 (58.5\%) are hard-to-cover. This highlights the prevalence of these challenges.}

There are several reasons which might contribute to the inability of the state-of-the-art techniques to handle branches with complex objects or inter-procedural dependencies.

First, during the test generation process, existing techniques~\cite{pynguin,codamosa} normally try to construct minimal or over-simplified objects, without taking into account the real usage scenarios involving specific attributes. Second, for the inter-procedural dependencies, state-of-the-art techniques~\cite{pynguin,codamosa} are not able to recursively analyze how dependent methods interact with variables from the target method, thus failing to interpret the concrete branch constraints. 
Third, existing LLM-based techniques~\cite{codamosa,chattester} typically provide either unnecessary or insufficient contexts for LLMs to generate tests. 
\coda{} provides the whole module as the context, where many parts of code are irrelevant to the target method and thus introduce noise to LLMs.
\tester{} provides the class declaration, constructor signatures, relevant fields, and the target method as the context, which misses the broader context dependent to the target method.

Based on these potential reasons behind the unsatisfactory performance of existing techniques, we propose \tool, an LLM-based test generation technique leveraging program-analysis-enhanced prompting. \tool{} will 1) conduct \del{backward method-invocation analysis}\ins{object construction analysis} to learn real usage scenarios of objects and overcome the difficulty of constructing complex objects, 2) conduct \del{forward method-invocation analysis}\ins{branch dependency analysis} to understand the behavior of complicated inter-procedural dependencies, and 3) supply relevant information to LLMs for test generation, avoiding imprecise context.

\section{Approach}
\label{sec:approach}

\begin{figure}[t]
  \centering
  \includegraphics[width=0.7\linewidth]{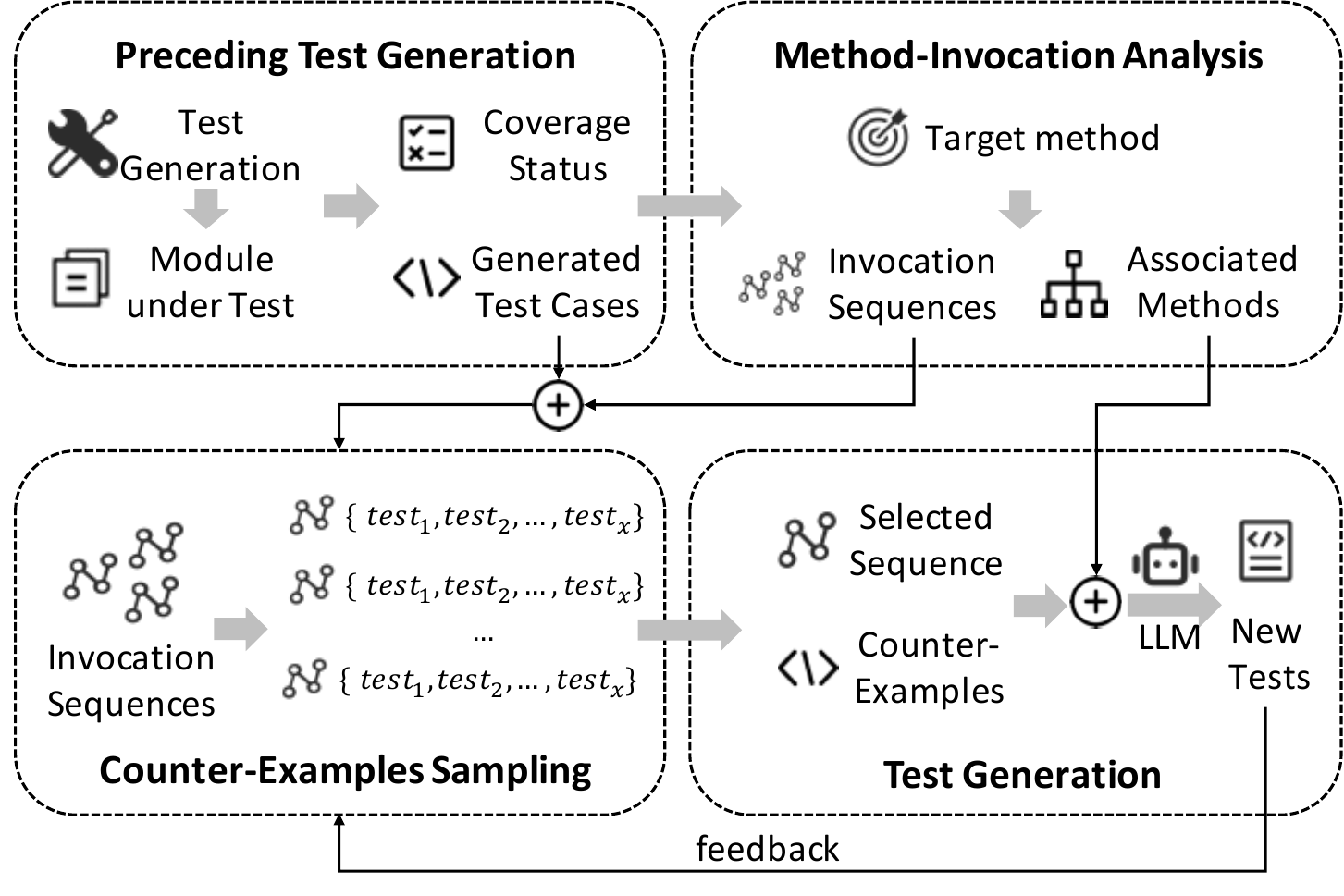}
  \caption{Overview of \tool{}}
  \label{fig:overview}
\end{figure}

\begin{figure}[t]
  \centering
  \includegraphics[width=0.45\linewidth]{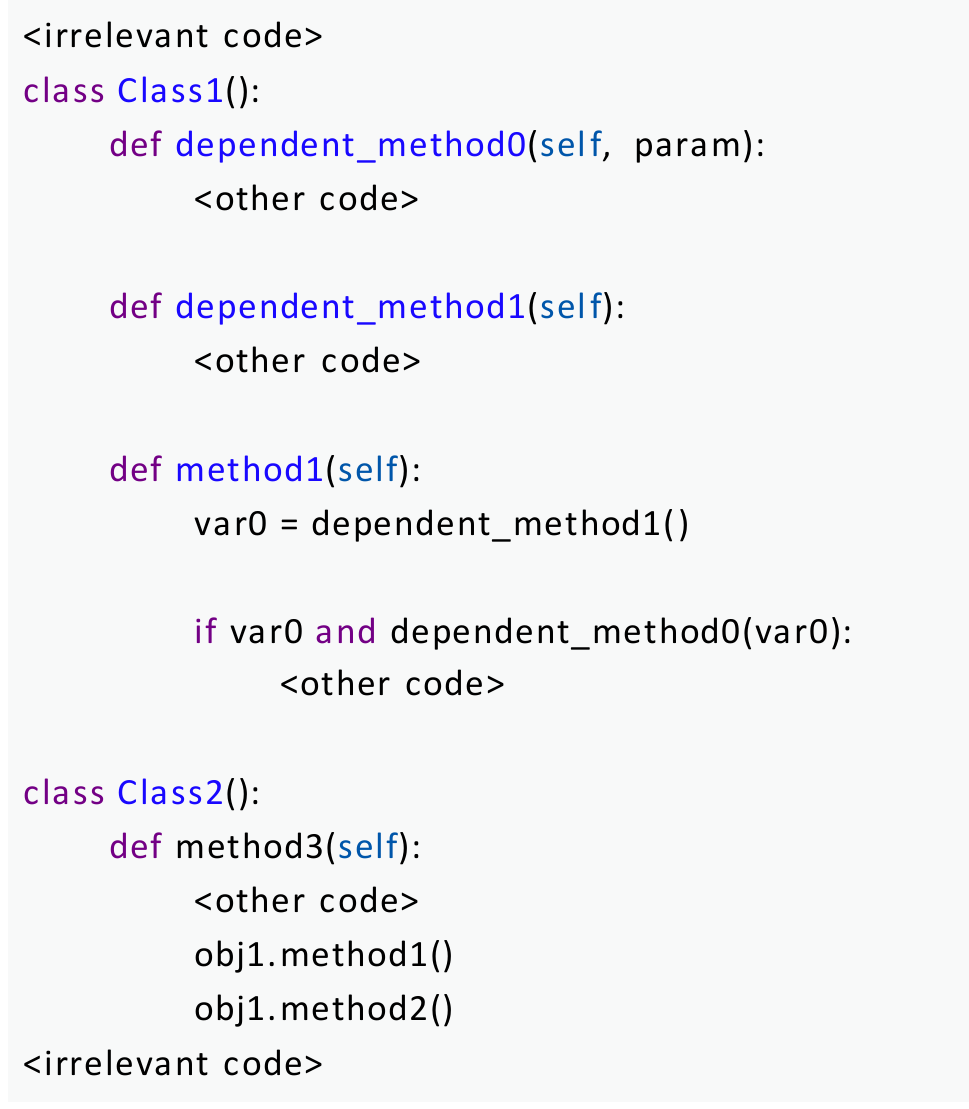}
  \caption{Illustrative Example}
  \label{fig:illustrative_example}
\end{figure}

The overall workflow of \tool{} is illustrated in Figure~\ref{fig:overview}. 
First, \tool{} leverages existing test generation tools to generate a set of tests, which can reach those easy-to-cover branches. 
For ease of presentation, we call this step \textit{preceding test generation}.
Then, for the methods containing the branches that are still not covered, \tool{} performs two types of program analyses\del{, including backward method-invocation analysis to retrieve the method invocation sequences for complex object construction and forward method-invocation analysis to retrieve all associated methods for inter-procedural dependencies}\ins{: 1) object construction analysis: This analysis traces backward from the target method to identify sequences of method calls that lead to the construction of objects used as parameters. By capturing real usage scenarios, this analysis helps the LLM generate tests that construct valid objects with the necessary attributes; 2) branch dependency analysis: This analysis explores forward from the branch conditions of the target method to identify all methods that influence the outcome of the conditions. By providing precise contextual information about these dependencies, this analysis enables the LLM to generate tests that effectively cover hard-to-reach branches}.
To improve the testing efficiency, \tool{} samples existing generated tests as counter-examples to guide LLMs to generate different tests, as they are shown ineffective to cover specific branches.
These counter-examples, together with the program analysis results, are incorporated into the prompt for LLMs to generate new tests. The new tests will be executed and added into existing tests for future iterations. 
In the following, we will introduce \tech{} in detail by using the example shown in Figure~\ref{fig:illustrative_example} for facilitating illustration.

\subsection{Preceding Test Generation with Existing Tools}
\label{sec:preceding_test_generation}
As our goal is to reach those hard-to-cover branches, \tech{} uses existing test generation tools for the initial testing task (for those easy-to-cover branches).
This helps improve the cost-effectiveness of the overall testing process due to the cost of utilizing LLMs.
Specifically, for a project under test, \tech{} runs an existing test generation tool until it is unable to cover new branches within a pre-defined timeframe.
\ins{
Following previous work~\cite{codamosa}, we conducted a pilot study on a small benchmark and then set the timeframe to 2 minutes based on the observed results in \tech{}.
}
When executing these generated tests, their corresponding coverage information is also recorded. 
\del{It is worth noting that \tool{} is in theory language-independent and not restricted to specific tools for generating initial tests.}
In this section, we mainly describe the idea behind \tool{}, and the implementation details will be presented in Section~\ref{sec:implementation}.

\subsection{Method-Invocation Analyses}

\tool{} first collects the methods containing branches not covered by the preceding testing process, and then performs \del{backward and forward method-invocation analyses}\ins{method invocation analyses} on these methods to gather information needed for handling complex object construction and inter-procedural dependencies. 
This process is done in ascending order of coverage achieved for the methods, as we assume that methods with the lower coverage are more difficult to tackle and there is larger room for coverage improvement. 
Unlike existing LLM-based techniques (e.g., \coda{}~\cite{codamosa}) that try to incorporate as much source code from the module as possible into prompts, we will only feed LLMs with the methods retrieved from method-invocation analyses as the context in order to reduce noise introduced by irrelevant code and alleviate computational burden of LLMs.

\subsubsection{\del{Backward Analysis}\ins{Object Construction Analysis}}
\label{sec:backward}
The goal of the \del{backward analysis}\ins{object construction analysis} is to extract method-invocation sequences ending with the target method. 
These sequences represent diverse real usage scenarios of the target method in the project and have a high chance to capture the whole process of constructing objects used in the target method. For example, a method $v_0$ may create an object \textit{o} and invoke the target method \textit{v} with \textit{o} as an argument. A \del{backward method-invocation analysis}\ins{object construction analysis} could collect the invocation sequence $v_0 \rightarrow $\textit{v} and trace back to the object construction process. 

To achieve this goal, \tool{} first constructs a method call graph within the module hosting the target method, and then extracts all paths terminating at the target method as the method invocation sequences.

\begin{figure}[t]
  \centering
  \includegraphics[width=0.8\linewidth]{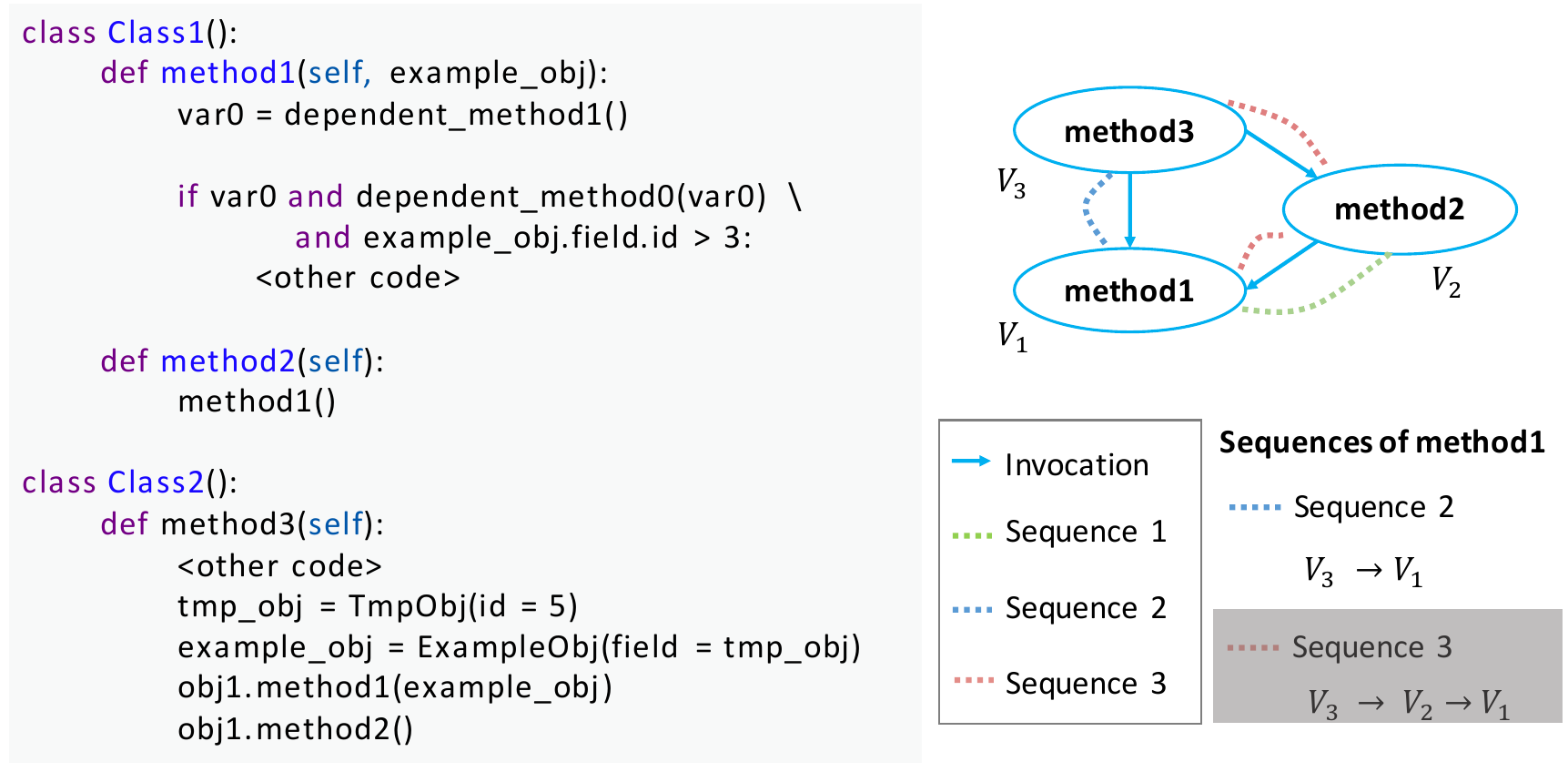}
  \caption{\ins{Example of object construction analysis}}
  \label{fig:call_graph_construction}
  \vspace{-1.5mm}
\end{figure}

\ul{Call Graph Construction.} \del{A directed call graph is constructed to represent the method invocation relationships among methods in the module hosting the target method.}\ins{\tool{} constructs a directed call graph to model method invocation relationships within the module containing the target method.} The nodes in the graph represent the methods in the module\del{ extracted by traversing the abstract syntax tree (AST)}, and the edges represent invocation relationships between pairs of methods.  
Figure~\ref{fig:call_graph_construction} illustrates a simplified version of the code shown in Figure~\ref{fig:illustrative_example} and the constructed call graph.
Given the simplified code snippet, the call graph will contain nodes \texttt{method1} (denoted by $v_1$) and \texttt{method2} (denoted by $v_2$) from \texttt{Class1}, and \texttt{method3}  (denoted by $v_3$) from \texttt{Class2}. Clearly, $v_1$ is called by $v_2$ and $v_3$, while $v_2$ is only called by $v_3$.

\del{For each extracted method, \tool{} also records the associated classes (the parent class and classes utilized within the method), as well as their declarations and constructors. This information is important because extracted methods might belong to or interact with variables/methods from multiple different classes. Understanding where these elements come from is critical for LLMs to construct valid objects.}

\ul{Sequence Extraction.} With the directed call graph,  \tool{} proceeds to extract sequences of method invocations for the target method. 
Specifically, \del{\tool{} identifies all the entry nodes from the graph and employs the Depth-First Search (DFS) strategy to gather all the paths originating from entry nodes and terminating at the target method.}\ins{Depth-First Search (DFS) is used to identify all paths starting from entry nodes and terminating at the target method.} If a cycle is encountered during path extraction, the traversal of that path will stop to prevent creating infinite paths. \del{It is worth noting that for all public methods inside a class, we consider each method itself as a complete path whose entry method is also the target method.}
Going back to the example in Figure~\ref{fig:call_graph_construction}, for the target method $v_1$, three paths will be extracted: $p_1 = \{v_3 \rightarrow v_2 \rightarrow v_1\}$, $p_2 = \{v_3 \rightarrow v_1\}$ and $p_3 = \{v_1\}$. \del{As can be seen here, there might be multiple call paths from the same entry method to the target method (e.g., $p_1$ and $p_2$).}

\del{To alleviate computational load, \tool{} employs a straightforward filtering strategy to condense the sequences with the same entry method to the target method. 
That is, only the shortest path is kept.}\ins{To alleviate computational load, only the shortest path from each entry point to the target method is retained.}
This strategy is used as such shorter and more direct call paths are easier for LLMs to understand and learn from.
In our example, the path $p_2$ is selected (that is, $p_1$ is filtered out).

\subsubsection{\del{Forward Analysis}\ins{Branch Dependency Analysis}} 
\begin{figure}[t]
  \centering
  \includegraphics[width=0.65\linewidth]{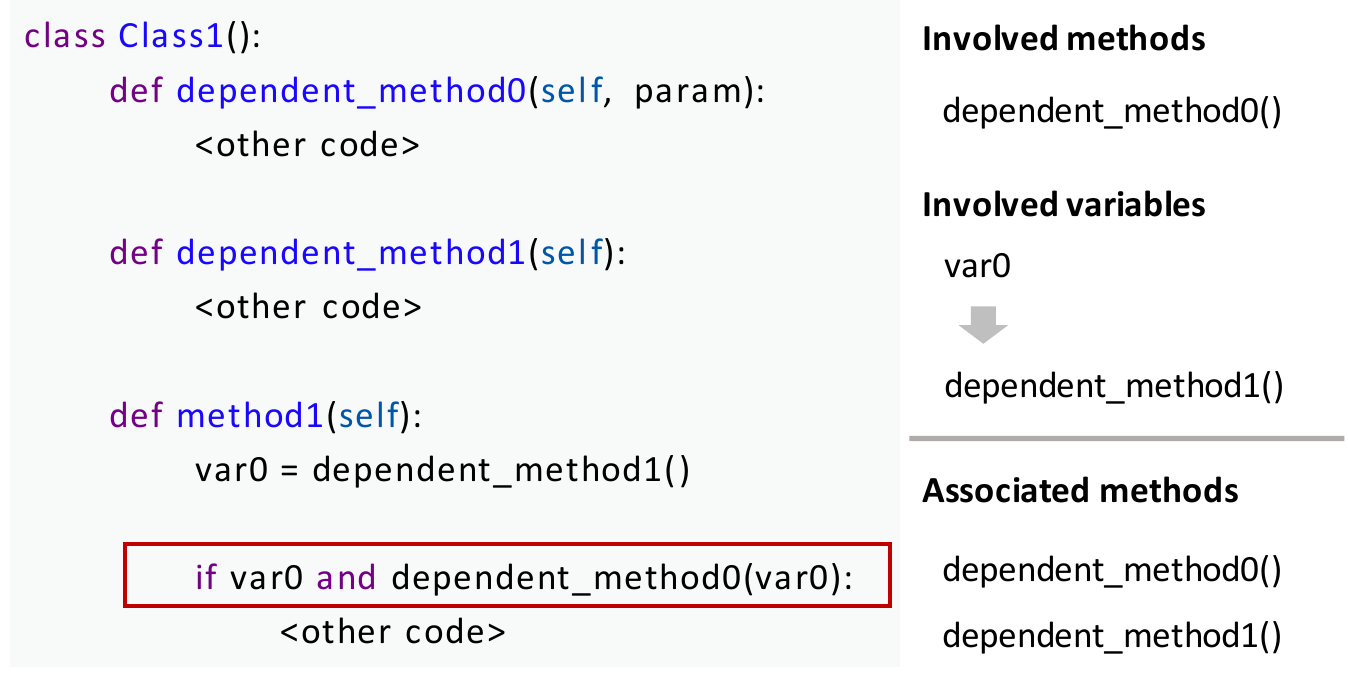}
  \caption{Example of \del{forward method-invocation analysis}\ins{branch dependency analysis}}
  \label{fig:branch_example}
\end{figure}

As illustrated in Section~\ref{sec:motivation}, it is challenging for test generation techniques to reach certain branches when the outcome of the branch condition depends on other invoked methods. To address it, \tech{} performs a \del{forward method-invocation analysis}\ins{branch dependency analysis} to recursively collect all methods that might be relevant to determining which branch to execute. The extracted methods provide rich information for interpreting what really entails in branch conditions. 
Below, we describe the \del{forward analysis} process in detail with the example presented in Figure~\ref{fig:branch_example}. 

To extract the set of associated methods $\mathcal{B}_v$ for a target method $v_1$ (\texttt{method1} in Figure~\ref{fig:branch_example}), \tech{} first identifies branch conditions and captures all variables (\texttt{var0}) and methods (\texttt{dependent\_method0}) that these condition expressions contain. 
All involved methods (\texttt{dependent\_method0}) are added to $\mathcal{B}_v$. 
For each involved variable, \tech{} recursively extracts relevant assignment statements. In our example, the value of \texttt{var0} is determined by \texttt{dependent\_method1}. Therefore, \tech{} considers \texttt{dependent\_method1} is relevant to \texttt{var0} and adds the involved methods (\texttt{dependent\_method1}) to $\mathcal{B}_v$.
In this way, all methods that can impact the outcome of branch conditions are retrieved and a precise search scope is identified to understand the behavior of condition expressions.  

\ins{
Note that \tech{} does not always use both branch dependency analysis and object construction analysis simultaneously. Specifically, it applies these analyses selectively based on the characteristics of the target method. Branch dependency analysis is used when the target method involves intricate inter-procedural dependencies in branch conditions, and object construction analyses is used when the target method involves complex objects in branch conditions. If the target method involves both complex objects and inter-procedural dependencies, \tech{} applies both analyses to address these challenges comprehensively.
For the simpler cases where the target method does not involve complex objects and inter-procedural dependencies, \tech{} provides the method under test itself as its context to the LLM, like the existing LLM-based approaches (such as \llm{}). This ensures that \tech{} remains effective for such cases.
}

\subsection{Counter-Examples Sampling}
\label{sec:few_shot_selection}

With the preceding testing process described in Section~\ref{sec:preceding_test_generation}, we have acquired a set of tests that fail to reach specific hard-to-cover branches. These existing tests can be used as counter-examples to guide LLMs to generate divergent tests, further enhancing the overall test efficiency. However, incorporating all existing generated tests into the prompt is impractical, which will not only impose significant computational burden on LLMs but also easily exceed the input length limit of the prompt.  
To mitigate this, \tech{} samples a diverse set of counter-example tests, striking a balance between effectiveness and efficiency.
As our goal is to reach hard-to-cover branches, it is essential to let LLM understand which branches have already been covered by existing tests. Meanwhile, as mentioned before, we want to have a small (preferably minimal) set of tests as counter-examples. 

To achieve this, \tech{} employs a coverage-based approach.
For each method invocation sequence of a target method $v$ obtained from the \del{backward analysis}\ins{object construction analysis}, \tech{} collects all the existing tests that invoke the first method of the sequence (denoted by $v_0$) as the candidate tests (denoted by $\mathcal{T}_{v_0}$). We only consider the first method of the sequence as it will in the end propagate to the target method through the chain of method calls. 
From $\mathcal{T}_{v_0}$, \tech{} first selects the test that achieves the highest coverage for the target method $v$, then picks the next tests that can achieve the highest incremental coverage. This is an iterative process until no more test can increase the coverage. In this way, the minimal set of tests can be acquired to reach all already-covered branches. 
\del{As can be seen,}\ins{Note that} the counter-examples are constructed for each method invocation sequence, this is because in the later stage of test generation, each time we will only feed one method invocation sequence to the LLM to alleviate the computational load and avoid exceeding the input length limit.

\subsection{Feedback-based Test Generation Process with LLMs}
\tool{} integrates program analysis results and counter-examples into prompts for the LLM to generate new tests. Previous studies have shown large performance improvement for LLMs through Chain-of-Thought (CoT)~\cite{cot,cot2}. Therefore, we also adopt a typical CoT strategy and divide the process into two stages. 
Note that \tech{} can also adopt different CoT techniques, as the contribution of \tech{} lies in improving the prompting approach instead of designing CoT techniques.

\begin{figure}[t]
  \centering
  \includegraphics[width=0.8\linewidth]{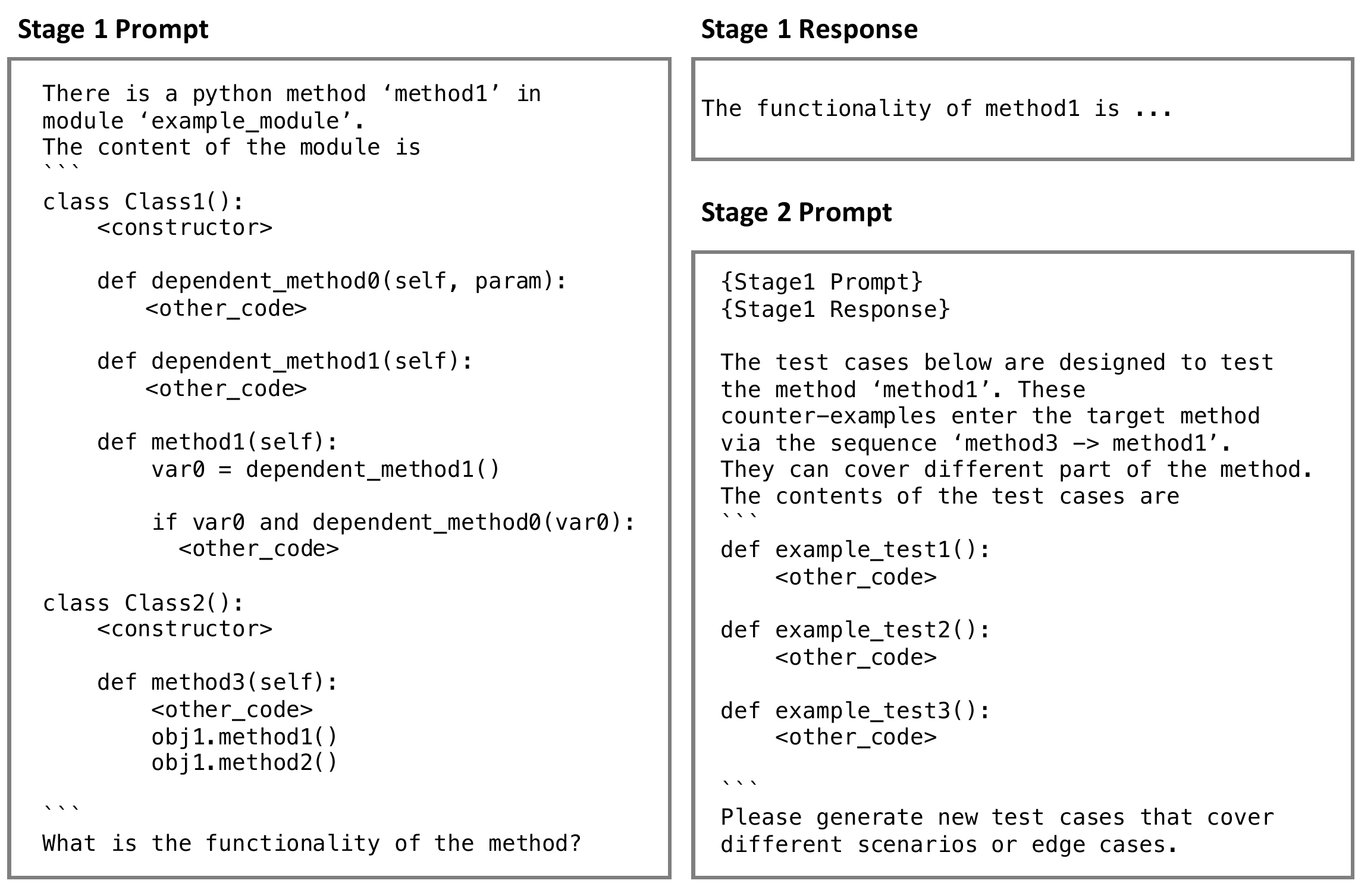}
  \caption{Example of prompt construction
  }
  \label{fig:prompt}
\end{figure}

In the first stage, \tool{} constructs the context of the target method $v$ with two sources: 1) methods in a selected method-invocation sequence from \del{backward analysis}\ins{object construction analysis}, and 2) the associated methods from \del{forward analysis}\ins{branch dependency analysis}.
For each method, \tool{} incorporates the declarations and constructors of their associated classes and the content of the method itself as the context information.
To select the method-invocation sequence for $v$ in each iteration, 
\tool{} first uses the sequence \texttt{$p = \{v\}$} (i.e., the entry method is $v$ itself), as this is the most straightforward sequence for the LLM to invoke and understand the semantics of $v$. 
If this sequence does not exist (e.g., the target method is not a public method), or the generated tests with it in the second stage do not cover new branches (note that our testing process is iterative with feedback),
\tool{} randomly selects an unused sequence. 
The context information is then fed to the LLM for summarizing the functionality of $v$, enabling the LLM to gain \del{deep understandings of the method semantics}\ins{contextual awareness of method semantics}. 
The test generation process for $v$ terminates when no sequence is left. 
For example, given \textit{method1} in Figure~\ref{fig:illustrative_example}, if the selected sequence is $p = {v_3 \rightarrow v_1}$, the prompt will be constructed using \textit{method3} from the \del{backward analysis}\ins{object construction analysis}, \textit{dependent\_method0} and \textit{dependent\_method1} from the \del{forward analysis}\ins{branch dependency analysis}, and \textit{method1} itself, as shown in Figure~\ref{fig:prompt}.

In the second stage, \tech{} integrates the contents of the sampled counter-examples corresponding to the method invocation sequence into the prompt, along with a description: ``These counter-examples enter the target method via the selected sequence of method invocations. They can cover different parts of the method. Please generate new test cases that cover different scenarios or edge cases.
'', as shown in Figure~\ref{fig:prompt}.
Such a description can facilitate the understanding of the LLM on the intent of the generation, namely to cover different branches in the target method following the method invocation sequence. The LLM is then instructed to generate divergent tests.

The new test generated by the LLM is executed and added to the existing tests along with its coverage information.
The used counter-examples are removed from the candidates to ensure the diversity in future counter-example selection. This type of feedback helps \tool{} generate more tests with different behaviors.

\section{Evaluation}
To evaluate the performance of \tech{}, we formulate the following research questions (RQs).
\begin{itemize}
   \item \textbf{RQ1}: \textit{To what extent can \tech{} improve code coverage compared to the state-of-the-art techniques? } \del{The goal of \tech{} is to reach hard-to-cover branches, and this RQ verifies its ability in this regard. }
   
   \item \textbf{RQ2}: \textit{How does each of the main components in \tech{} affect the performance?} \del{This RQ leads to an ablation study to examine how \tech's main components (i.e., method-invocation analysis, counter-example sampling, coverage-based feedback) impact the performance.}
   
   \item \textbf{RQ3}: \textit{How does the adoption of preceding test generation tools and LLMs impact the performance of \tech{}?} \del{Our \tech{} uses existing tools for initial test generation, and utilizes an LLM to generate new tests. In this RQ, we examine how the adoption of existing tools and the choice of the LLM can impact the performance.}

   \ins{
      \item \textbf{RQ4}: \textit{How is the correctness of the unit tests generated by \tech{}?}
   }

   \insminor{
      \item \textbf{RQ5}: \textit{How generalizable is \tech{}?}
   }

\end{itemize}

\subsection{Experimental Setup}
\subsubsection{Implementation of \tech{}}
\label{sec:implementation}

While \tech{} is not limited to specific programming languages, in this experiment, we target test generation for Python programs\insminor{ in RQ1-RQ4}, given the popularity of Python and the existing Python benchmarks~\cite{codamosa}. \insminor{Additionally, we adapted \tool{} to Java to evaluate its generalizability in RQ5.}
\delminor{\tech{} is also developed in Python 3.10 with over 5k lines of code. }
\del{We conducted static program analysis using Python's \textit{AST} module and performed coverage analysis with \textit{coverage.py}~\cite{coverage.py}.}

\ins{\insminor{Regarding the Python version of \tool{}, to}\delminor{To} construct the call graph, we rely on PyCG~\cite{pycg}, a state-of-the-art call graph generator for Python. 
PyCG provides a robust foundation for identifying method invocation relationships within the code. 
PyCG constructs an assignment graph that tracks the assignment relations between program identifiers, such as variables and classes. Furthermore, it uses namespace and scope resolution to distinguish between methods with the same name in different classes.
To better align with \tool{}'s requirements for sequence extraction and prompt construction, we map the nodes in the PyCG-extracted call graph back to the corresponding AST nodes parsed using Python's \textit{AST} module. 
Since TELPA focuses on generating tests for the target module, we primarily consider method invocations within the module under test. External functions (e.g., those from imported libraries) are not further analyzed for sequence extraction, as their internal implementations are outside the scope of the target module.
}
For coverage analysis, \tool{} uses \textit{coverage.py}~\cite{coverage.py}, a widely-used tool for measuring code coverage in Python programs.
The LLM \textit{Phind-CodeLlama-34B-v2}~\cite{phind} from Hugging Face~\cite{huggingface} is employed for test generation by default, as it is one of the most effective open-source code LLMs according to the LLM leaderboard maintained by Hugging Face~\cite{leaderboard}. \textit{PyTorch 1.13}~\cite{torch}, \textit{Hugging Face Transformers 4.35.2}~\cite{transformers}, and \textit{fastchat 0.2.30}~\cite{fastchat} are used to run the LLM locally. Following the existing work~\cite{fuzz4all,whitefox}, we set the temperature to 0, enabling \tool{} to obtain more deterministic results from the LLM.

To reduce the resource consumption, \tech{} uses state-of-the-art techniques for initial test generation. In our study, we adopted two test generation tools separately: a state-of-the-art SBST \del{technique}\ins{tool} \sbst{}~\cite{pynguin} and a state-of-the-art \del{LLM-based}\ins{LLM-enhanced} technique \coda{}~\cite{codamosa}.  
That is, we constructed two instances of \tool{}: \sbst-based \pynguinbase{} and \coda-based \codamosabase{}. The adoption of these two different types of preceding test generation techniques ensures the conclusion generalizability. More information regarding \sbst{} and \coda{} will be given in Section~\ref{sec:rq1_baselines}.

\insminor{
Specifically, to answer RQ5, we adapted \tool{} to Java.
Note that although \tech{} is in theory language-independent, it still requires engineering efforts as the analysis and the parsing of test cases are highly specific to the chosen frameworks.
Thus we made the engineering effort to extend TELPA to support Java projects.
Specifically, we adapted TELPA's program analysis components to work with Java code by incorporating JavaParser~\cite{javaparser}, allowing to effectively parse Java code. 
We also implemented the test execution process using the JUnit framework (the standard framework for unit testing in Java) and incorporated Jacoco~\cite{jacoco} for code coverage analysis (a widely-used tool for measuring Java code coverage). 
Additionally, we modified the processing of the LLM’s output to extract Java tests and address potential issues such as missing external libraries. 
}

\insminor{
Similarly, to reduce resource consumption, \tool{} leverages the state-of-the-art search-based test generation tool for Java (i.e., \evo{}~\cite{evo}) for preceding test generation.
Specifically, \evo{} is used to generate initial tests, when \evo{} no longer improves coverage within a set time window, \tech{} is activated, takes over the test generation process entirely and leverages \evo{}-generated tests as counter-examples.
The LLM is used in the same way as the Python version of \tool{}.

}

As mentioned before, %
\tech{} is activated when the preceding tool fails to increase coverage within a pre-defined timeframe.
As \del{the }previous work~\cite{codamosa}, we conducted a pilot study on a small test benchmark to optimize the performance of \tool{}. The timeframe is set to 2 minutes based on the observed results, namely \tech{} will take over the test generation task if \delminor{\sbst{} or \coda{}}\insminor{\sbst{}, \coda{} or \evo{}} cannot increase the coverage for two minutes. %
This choice is further discussed in Section~\ref{sec:threats}.

\subsubsection{Baselines for RQ1\ins{ and RQ4}}
\label{sec:rq1_baselines}
To answer RQ1\ins{ and RQ4}, we compared \tech{} with three baselines: \textbf{\sbst{}}, \textbf{\coda{}}, and \textbf{\tester{}}.

\sbst{}~\cite{pynguin} is the state-of-the-art SBST \del{technique}\ins{tool} for \del{dynamically typed programming languages like }Python.
It initiates the testing process by taking the Python code under test as input, and employs \del{one of }search-based algorithms (including MIO~\cite{mio1, mio2}, MOSA~\cite{bothmosa}, and DynaMOSA~\cite{bothmosa}) to generate new tests by mutating values and statements of a seed test. 
Here, we used \textit{Pynguin 0.19.0} with DynaMOSA as this algorithm has been demonstrated to be the most effective one among all supported algorithms in \sbst{}~\cite{pynguin}.

\coda{}~\cite{codamosa} is the state-of-the-art LLM-based test generation tool, which was proposed for improving SBST techniques.
It prompts LLMs to generate seed tests for \sbst{} in order to help SBST escape the local optimum.
More specifically, when \sbst{} falls into the local optimum, \coda{} switches to invoke LLMs to generate new seed tests by incorporating as much source code of the module under test into the prompt as possible, and then switches back to \sbst{} for running on these new seed tests, until the given testing budget is reached. 
\ins{
Specifically, we consider \coda{} as an integral tool integrating both \sbst{} and the LLM, and then incorporate it into the preceding test generation component of \tech{}.
In the case of using CODAMOSA as the preceding test generation tool, when the switch (between Pynguin and the LLM) within CODAMOSA does not improve the coverage any more within the given timeframe (exceeding the switching time used within \coda{}), \tech{} makes \coda{} terminate and then activates its program-analysis-enhanced LLM-based test generation component for further coverage improvement.
}
\coda{} originally employed \textit{Codex} as the LLM for generating seed tests. However, as the Codex service has been shut down~\cite{codex_shutdown}, we employed \textit{Phind-CodeLlama-34B-v2} as the LLM for \coda{} to have a fair comparison with our technique.
\ins{We call it \codarevised{}.}

\tester{}~\cite{chattester} is another state-of-the-art LLM-based test generation tool, which first extracts context information for ChatGPT to generate tests and then designs post-processing strategies to fix invalid tests.
The main contribution of \tech{} is the novel prompting method. 
Unlike \del{\coda{}}\ins{\codarevised{}} that incorporates as much source code of the module under test as possible for prompting, or \tester{} that incorporates insufficient context for prompting, \tech{} feeds the code information highly relevant to branch constraints to LLMs.
\ins{
However, \codarevised{} and TELPA differ in both their LLM invocation workflows and their prompting methods. Specifically, \codarevised{} iteratively switches between the search-based tool \sbst{} and the LLM, while once \tech{} switches, it does not revert back to the preceding tool. That is, \tech{} continues to use the LLM with its program-analysis-enhanced prompting for subsequent test generation.
Additionally, \codarevised{} uses a simple prompting method by incorporating the entire module’s source code as context for the LLM. \tech{} introduces a novel prompting method that leverages program analysis to provide precise and relevant context for the LLM.
}
\del{The}\ins{Therefore, the} comparison with the \del{LLM-based \coda{}}\ins{LLM-enhanced \codarevised{}} cannot clearly demonstrate the impact of this new prompting method due to the different LLM invocation workflows.
Therefore, we used \tester{} as another LLM-based baseline.
To highlight the effectiveness of advanced prompting methods for test generation, we directly replaced our novel prompting method with the one used by \tester{} and removed the post-processing strategy from \tester{}. 
Note that this component is orthogonal to the test generation process and can be also applied to \tool{}. 
\ins{
That is, we activated the prompting method of CHATTESTER for test generation when the preceding test generation tool reaches a coverage plateau, same as the workflow of TELPA.
}
\tester{} originally employed ChatGPT as the LLM for generating tests, we replaced it with \textit{Phind-CodeLlama-34B-v2} for fair comparison. 
\ins{We call it \llmrevised{}.}
Since both \sbst{} and \del{\coda{}}\ins{\codarevised{}} are used as preceding test generation tools in our study, we also created two instances of \llm{} for comparisons: \sbst-based \textbf{\del{\llmsbst{}}\ins{\llmsbstrevised{}}} and \coda-based \textbf{\del{\llmcoda{}}\ins{\llmcodarevised{}}}.

\del{To sum up, we have two sets of comparisons: 1) \pynguinbase{} vs. Pynguin vs. \llmsbstrevised{} and 2) \codamosabase{} vs. CODAMOSA vs. \llmcodarevised{}.}
\ins{To sum up, we have two sets of comparisons: 1) \pynguinbase{} vs. \sbst{} vs. \llmsbstrevised{} and 2) \codamosabase{} vs. \codarevised{} vs. \llmcodarevised{}.}
These comparisons aim to investigate whether \tech{} can improve the existing test generation tools (\sbst{} and \del{\coda{}}\ins{\codarevised{}}) and examine whether our novel prompting method can outperform the state-of-the-art prompting method proposed by \tester{} regardless of the preceding test generation tools.

\subsubsection{Variants for RQ2}
\label{sec:rq2_baselines}
To answer RQ2, we constructed \del{six}\ins{seven} variants of \tech{} for the ablation experiment to investigate the contribution of each component in \tech{} :

\begin{itemize}

    \item \textbf{\nocall{}} that removes the \del{backward method-invocation analysis}\ins{object construction analysis} from \tech{};
    \item \textbf{\nobranch{}} that removes the \del{forward method-invocation analysis}\ins{branch dependency analysis} from \tech{};
    \item \textbf{\noshot{}} that removes the counter-example guidance from \tech{};
    \item \textbf{\randomshot{}} that randomly selects the same number of counter-examples instead of coverage-based sampling in \tech{};
    \item \textbf{\nofeedback{}} that removes the feedback process from \tech{}. Specifically, the newly generated tests by the LLM, along with their coverage information, will not be added to the pool of existing tests by \nofeedback{}. 
    \item \textbf{\nofiltering{}} that removes the path filtering process from \tech{}. Specifically, instead of condensing the sequences with the same entry and ending method, we keep all sequences.
    \ins{
    \item \textbf{\nocot{}} that removes the functionality summarization process from \tech{}.
    }
   
\end{itemize}

Similar to RQ1, we constructed two sets of variants based on the preceding test generation tool: \sbst-based variants (\nocallsbst{}, \nobranchsbst{}, \noshotsbst{}, \randomshotsbst{}, \nofeedbacksbst{}, \nofilteringsbst{}\ins{, \nocotsbst{}}); and \coda-based variants (\nocallcoda{}, \nobranchcoda{}, \noshotcoda{}, \randomshotcoda{}, \nofeedbackcoda{}, \nofilteringcoda{}\ins{, \nocotcoda{}}).

\subsubsection{Variants for RQ3}
\tech{} is built on top of the adopted preceding test generation tools and LLM. To better understand how this design choice will impact the performance of \tech{}, in this RQ, we construct another two variants. 
In the first variant, named \textbf{\techzero{}}, no preceding test generation tool is used. That is, \tech{} is activated from the very beginning. 
In the second variant, we consider a different LLM.
The default LLM in \tech{} is \textit{Phind-CodeLlama-34B-v2}, which is relatively large-scale and may require a lot of resources to run.
To make \tech{} more practical, we also investigated the effectiveness of \tech{} when a relatively small-scale LLM is adopted. More specifically, we employed \textit{DeepSeek-Coder-6.7B-instruct}~\cite{deepseek}, which is one of the most effective LLMs among the set of LLMs with comparable (small) scales on HuggingFace\ins{ according to the LLM leaderboard maintained by Hugging Face~\cite{leaderboard}}.
Accordingly, we call the variant \deepseekbase{}. Similar to previous RQs, \sbst{}-based and \del{\coda{}}\ins{\codarevised{}}-based variants are created and named as \textbf{\deepseeksbst{}} and \textbf{\deepseekcoda{}}.
Furthermore, the commercial closed-source LLMs developed by OpenAI, such as ChatGPT and GPT-4, have shown superior performance on various benchmarks. Therefore, in addition to open-source LLMs, we also conducted an experiment to understand the effectiveness of \tool{} when using the more advanced GPT-4. 
Due to the high cost of invoking GPT-4's APIs, we only constructed a Pynguin-based variant, i.e., \gptsbst{}.

\subsubsection{\insminor{Baseline for RQ5}}
\insminor{
To answer RQ5, we adapted \tool{} to Java and constructed \evosuitebase{}, which leverages the state-of-the-art search-based test generation tool for Java (i.e., \textbf{\evo{}}~\cite{evo}) for preceding test generation and compared it with \evo{}. }

\insminor{
\evo{} is one of the most prominent search-based techniques. It employs evolutionary algorithms to generate unit tests aimed at maximizing code coverage. Additionally, it integrates advanced program analysis techniques to further enhance its effectiveness, including leveraging a constant pool for generating test inputs and applying testability transformations to improve guidance during the test generation process, etc.
}

\subsubsection{Benchmark, Metrics, and Environment}
\insminor{To answer RQ1-RQ4, following}\delminor{Following} the existing work~\cite{codamosa}, we evaluated \tech{} on the widely-studied benchmark, which consists of 486 modules from 27 Python projects. 
The whole benchmark contains 42,897 source lines of code and 4,518 methods in total. 
\ins{Since \tech{} focuses on improving coverage for hard-to-cover branches, we also analyzed the branches within the benchmark. The benchmark contains a total of 6,559 branches, of which 3,979 are classified as hard-to-cover. Specifically, 3,931 branches involve complex dependencies, while 256 branches are associated with complex objects.}

\insminor{
To answer RQ5, we further evaluated \tech{} on four real-world open-source projects using the JUnit framework~\cite{junit} in the widely-used Defects4J benchmark~\cite{d4j} due to the significant popularity of Java and JUnit  (i.e., Chart, Time, Lang, and Math). 
Unlike Python, Java is a statically typed language. Automated test case generation for dynamically typed languages often struggles with object creation, as obtaining detailed type information can be challenging. In contrast, Java’s static typing allows for more precise and efficient handling of type information, making it an ideal choice for evaluating the effectiveness of TELPA on static type systems.
Here, we excluded Closure since it does not use the JUnit framework and involves significant testing costs.
In total, the benchmark contains 9,391 branches, of which 5,494 are classified as hard-to-cover. 
Specifically, 2,453 branches involve complex dependencies, while 4,842 branches are associated with complex objects.
}

Following the existing work on test generation~\cite{pynguin,codamosa}, we adopted both branch coverage and line coverage as metrics to measure the effectiveness of a test generation technique.
\ins{Specifically, to evaluate the correctness of the generated tests cases in RQ4, we executed all generated test cases and measure the correctness using two metrics: (1) \textit{Syntax Correctness} refers to the percentage of generated tests that are syntactically valid. Specifically, we checked this by parsing the generated tests using Python's built-in ast module. Any test that passed parsing without errors was considered syntactically correct. (2) \textit{Execution Pass Rate} refers to the percentage of generated tests that executed successfully without runtime errors. Specifically, we ran the tests and recorded any runtime errors or failures during execution.
}
For each module and each technique in a project, we allocated 20 minutes for test generation. All the techniques kept running until the time budget ran out, even if the coverage did not grow before that. Note that the time costs spent on all components in \tech{} (including the preceding test generation, program analysis, test generation, and the coverage analysis) were included in this time budget.
\ins{
The 20-minute time budget was selected to ensure that all the studied techniques have sufficient time to explore the test space and demonstrate their potential as much as possible. 
While the previous studies set shorter time budgets for evaluation (e.g., 10 minutes for evaluating \sbst{} and \coda{}), a longer time budget is helpful to assess the sustained effectiveness of each technique over a sufficient period.
}

To avoid the influence of randomness, we repeated our experiments for \del{three}\ins{20} times considering the evaluation cost and used the average value of the final results. The experiments were conducted on a workstation with 128-core CPU, 504G memory, 4 NVIDIA A100 GPUs, and Ubuntu 20.04 OS.

We released our implementation and all experimental data at the project homepage~\cite{homepage} to facilitate replication, future research, and practical use.

\subsection{RQ1: Effectiveness of \tool{}}
\begin{table*}[t]
\setlength{\tabcolsep}{2.5pt}
\caption{Comparison among \tech{}, \llm{}, \sbst{} and \coda{} in terms of branch coverage on all branches}
\centering
\small
\label{tab:rq1_coverage}
\renewcommand{\arraystretch}{1.4}
\resizebox{0.99\linewidth}{!}
{
\begin{threeparttable}
\begin{tabular}{l|rrrr|rrrr}
\toprule
\textbf{Project}  & \textbf{\pynguinstall{}} & \textbf{\sbst{}} & \textbf{\llmsbstrevised{}} & \textbf{\pynguinbase{}} & \textbf{\codastallrevised{}} & \textbf{\codarevised{}} & \textbf{\llmcodarevised{}} & \textbf{\codamosabase{}} \\
\midrule
pysnooper           & 16.67\% & 18.07\% & 20.00\% & \textbf{31.23\%}  & 16.67\% & 18.25\% & 27.37\% & \textbf{30.70\%} \\
apimd               & 39.63\% & 43.18\% & 48.04\% & \textbf{81.50\%}  & 6.92\%  & 39.81\% & 51.78\% & \textbf{67.94\%} \\
blib2to3            & 23.55\% & 26.09\% & 25.18\% & \textbf{42.39\%}  & 19.59\% & 25.58\% & 30.66\% & \textbf{38.78\%} \\
codetiming          & 65.00\% & 75.00\% & 80.00\% & \textbf{95.00\%}  & 70.00\% & 70.00\% & 60.00\% & \textbf{95.00\%} \\
cookiecutter        & 45.37\% & 54.15\% & 63.90\% & \textbf{64.63\%}  & 45.37\% & 57.80\% & 52.44\% & \textbf{67.07\%} \\
dataclasses\_json   & 16.11\% & 16.37\% & 21.77\% & \textbf{27.96\%}  & 17.17\% & 17.61\% & 17.17\% & \textbf{28.94\%} \\
docstring\_parser   & 43.58\% & 53.77\% & 72.64\% & \textbf{85.85\%}  & 83.77\% & 86.04\% & 89.62\% & \textbf{91.13\%} \\
flutes              & 68.61\% & 77.22\% & 81.94\% & \textbf{84.17\%}  & 68.06\% & 76.67\% & 76.11\% & \textbf{82.50\%} \\
flutils             & 46.52\% & 47.41\% & 59.70\% & \textbf{78.89\%}  & 66.22\% & 66.74\% & 67.26\% & \textbf{78.00\%} \\
httpie              & 28.14\% & 29.79\% & 38.35\% & \textbf{45.52\%}  & 26.49\% & 27.11\% & 40.05\% & \textbf{44.90\%} \\
isort               & 94.00\% & 99.00\% & 98.00\% & \textbf{100.00\%} & 94.00\% & 98.00\% & 94.00\% & \textbf{98.00\%} \\
mimesis             & 76.97\% & 80.42\% & 80.25\% & \textbf{89.75\%}  & 68.57\% & 79.58\% & 83.45\% & \textbf{89.66\%} \\
py\_backwards       & 28.62\% & 42.00\% & 50.00\% & \textbf{63.69\%}  & 31.69\% & 36.00\% & 33.23\% & \textbf{60.31\%} \\
pymonet             & 62.67\% & 65.33\% & 68.00\% & \textbf{85.33\%}  & 61.33\% & 63.33\% & 71.56\% & \textbf{88.22\%} \\
pypara              & 44.13\% & 56.35\% & 47.14\% & \textbf{63.97\%}  & 18.41\% & 48.57\% & 34.13\% & \textbf{56.51\%} \\
semantic\_release   & 37.50\% & 38.33\% & 45.56\% & \textbf{60.28\%}  & 37.22\% & 37.22\% & 39.72\% & \textbf{60.00\%} \\
string\_utils       & 83.75\% & 85.62\% & 93.12\% & \textbf{97.97\%}  & 64.84\% & 80.62\% & 85.31\% & \textbf{87.97\%} \\
pytutils            & 35.95\% & 37.03\% & 50.95\% & \textbf{57.57\%}  & 41.08\% & 45.14\% & 43.92\% & \textbf{57.57\%} \\
sanic               & 44.56\% & 44.91\% & 59.65\% & \textbf{63.07\%}  & 46.23\% & 48.60\% & 51.58\% & \textbf{59.91\%} \\
sty                 & 87.14\% & 90.00\% & 91.43\% & \textbf{95.71\%}  & 90.00\% & 93.57\% & 89.29\% & \textbf{95.00\%} \\
thefuck             & 19.88\% & 20.71\% & 21.07\% & \textbf{50.60\%}  & 21.90\% & 23.45\% & 38.69\% & \textbf{54.17\%} \\
thonny              & 17.35\% & 36.63\% & 34.58\% & \textbf{38.31\%}  & 27.11\% & 36.51\% & 37.95\% & \textbf{43.13\%} \\
tornado             & 46.30\% & 47.89\% & 55.75\% & \textbf{65.36\%}  & 42.79\% & 44.05\% & 51.20\% & \textbf{65.61\%} \\
tqdm                & 15.00\% & 18.48\% & 40.45\% & \textbf{44.70\%}  & 35.30\% & 42.12\% & 38.33\% & \textbf{49.24\%} \\
typesystem          & 33.08\% & 37.50\% & 39.46\% & \textbf{50.87\%}  & 20.94\% & 56.09\% & 57.32\% & \textbf{72.54\%} \\
youtube\_dl         & 15.28\% & 16.97\% & 19.46\% & \textbf{25.03\%}  & 20.70\% & 23.57\% & 24.56\% & \textbf{29.01\%} \\
ansible             & 28.00\% & 29.86\% & 30.88\% & \textbf{38.02\%}  & 29.73\% & 31.38\% & 30.76\% & \textbf{37.78\%} \\
\midrule
Average Branch Cov. & 43.09\% & 47.71\% & 53.23\% & \textbf{63.98\%}  & 43.41\% & 50.87\% & 52.50\% & \textbf{64.06\%} \\

Average Line Cov.   & 62.36\%        & 64.72\%           & 70.20\%               & \textbf{75.54\%}      & 58.37\%         & 67.92\%            & 72.64\%                & \textbf{74.25\%}       \\
\bottomrule
\end{tabular}
\end{threeparttable}
}

\end{table*}

\begin{table*}[t]
\setlength{\tabcolsep}{2.5pt}
\caption{\ins{Comparison among \tech{}, \llmrevised{}, \sbst{} and \codarevised{} in terms of branch coverage on different types of branches}}
\centering
\small
\label{tab:rq1_hard}
\renewcommand{\arraystretch}{1.4}
\resizebox{0.99\linewidth}{!}
{
\begin{threeparttable}
\ins{
\begin{tabular}{lc|rrrr|rrrr}
\toprule
\multicolumn{2}{c|}{\textbf{Branch Type}}  & \textbf{\pynguinstall{}} & \textbf{\sbst{}} & \textbf{\llmsbstrevised{}} & \textbf{\pynguinbase{}} & \textbf{\codastallrevised{}} & \textbf{\codarevised{}} & \textbf{\llmcodarevised{}} & \textbf{\codamosabase{}} \\
\midrule
\multicolumn{2}{c|}{Easy-to-cover} & 64.50\% & 72.51\% & 77.17\% & 78.84\% & 73.10\% & 78.18\% & 80.65\% &  82.32\% \\
\midrule
\multicolumn{1}{c|}{\multirow{2}{*}{Hard-to-cover}}   & Branches with complex dependencies & 29.00\% & 31.16\% & 32.15\% & 39.10\% & 18.75\% & 20.17\% & 28.80\% & 34.90\% \\
\multicolumn{1}{c|}{}  & Branches with complex objects      & 52.73\% & 57.03\%  & 58.98\%  & 63.67\% & 39.06\% & 42.97\% & 53.90\%  & 58.59\%  \\
\cline{2-10}
\multicolumn{1}{c|}{}     &  \makecell{All hard-to-cover branches \\ (union of the above two types)}   & 29.20\% & 31.62\% & 32.67\% & 39.53\% & 19.13\% & 20.58\% & 29.10\% & 35.23\% \\
\bottomrule
\end{tabular}
}
\end{threeparttable}
}
\end{table*}

\ins{The goal of \tech{} is to reach hard-to-cover branches, and this RQ verifies its ability in this regard.}
Table~\ref{tab:rq1_coverage} presents the comparison results for \pynguinbase{} vs. \sbst{} vs. \del{\llmsbst{}}\ins{\llmsbstrevised{}} and \codamosabase{} vs. \del{\coda{}}\ins{\codarevised{}} vs. \del{\llmcoda{}}\ins{\llmcodarevised{}} on each project in terms of branch coverage. 
To gain a better understanding of the performance of the preceding test generation step, we also show the branch coverage achieved by \sbst{} and \del{\coda{}}\ins{\codarevised{}} right before switching to LLMs in ``\pynguinstall{}'' and ``\del{\codastall{}}\ins{\codastallrevised{}}'' columns\footnote{Here, ``bn'' represents ``bottleneck''.},  respectively. 
The last two rows present the average branch coverage and line coverage across all projects.
Due to the space limit, we do not show the line coverage for each project but instead include these detailed results in our replication package~\cite{homepage}.
\ins{To provide a clearer comparison of those techniques on hard-to-cover branches, Table~\ref{tab:rq1_hard} specifically shows the branch coverage achieved by each technique on different types of branches.}

\begin{figure}[t]   
  \centering     
  \begin{minipage}[t]{0.44\textwidth}
  \subfloat[\footnotesize Branch coverage trend of \pynguinbase{} and \sbst{}]{
        \includegraphics[width=\linewidth]{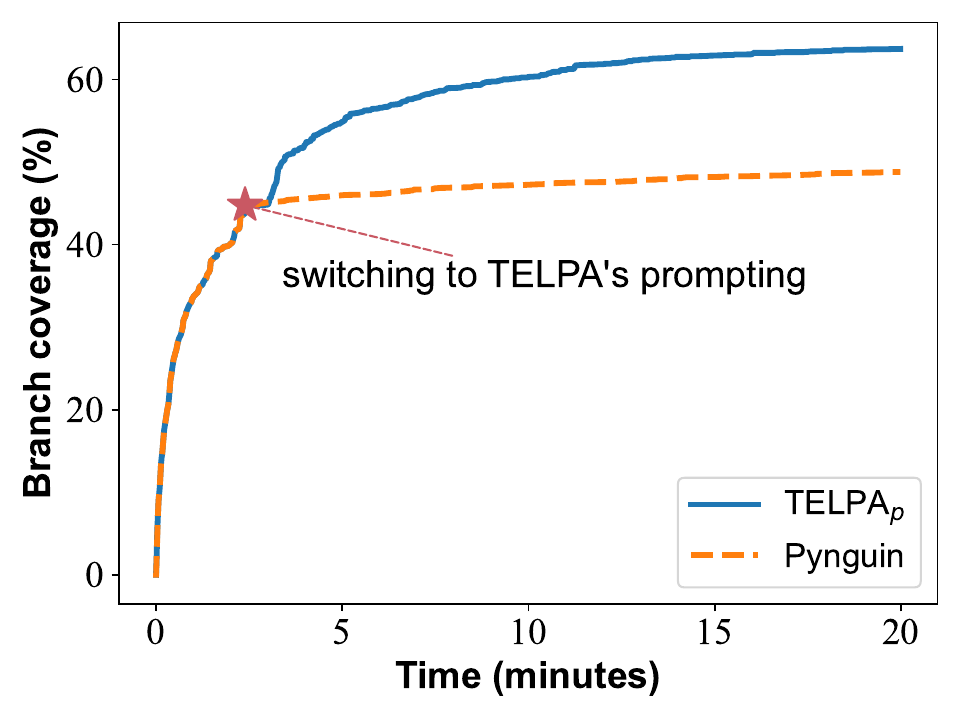}
        \label{fig:coverage_trend_coda}
        }
  \end{minipage}
  \begin{minipage}[t]{0.44\textwidth}
  \subfloat[\footnotesize Branch coverage trend of \codamosabase{} and \del{\coda{}}\ins{\codarevised{}}]{
      \includegraphics[width=\linewidth]{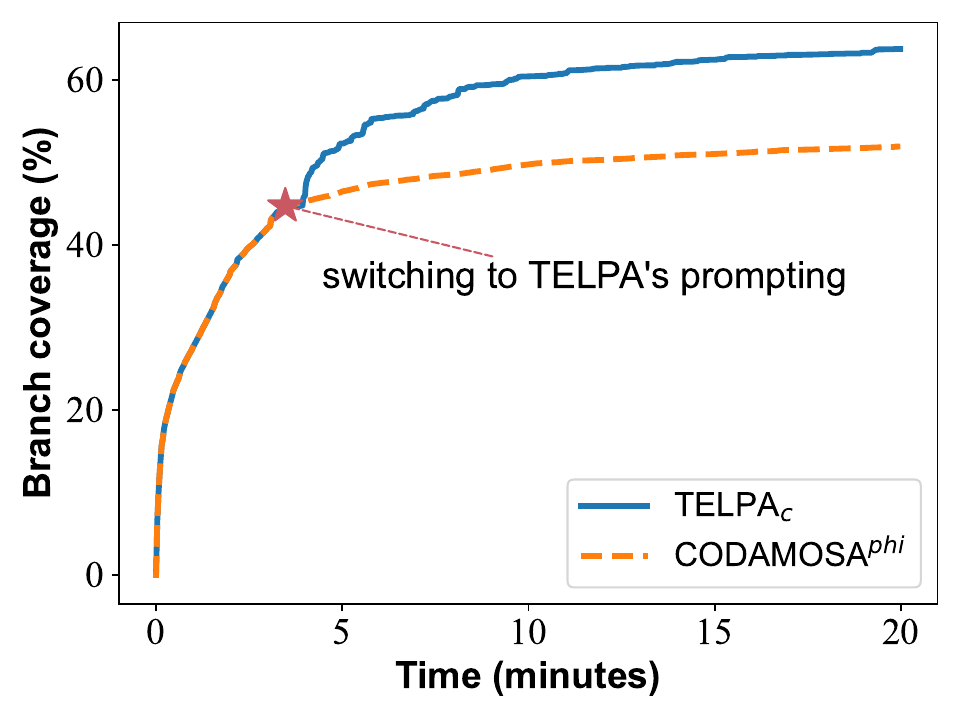}
        \label{fig:coverage_trend_coda}
    }
  \end{minipage}
  \caption{Branch coverage trend}
\label{fig:coverage_trend}
\end{figure}

To better understand the effectiveness of \pynguinbase{} and \codamosabase{}, we also measured the time spent on running \sbst{} and \del{\coda{}}\ins{\codarevised{}} before switching to LLMs.
On average across all projects, \sbst{} and \del{\coda{}}\ins{\codarevised{}} ran for 2.39 and 3.47 minutes, respectively, before switching.
That is, \tech{} is activated at an early stage of the entire testing process (given a time budget of 20 minutes) regardless of taking \sbst{} or \del{\coda{}}\ins{\codarevised{}} as the preceding test generation tool.
When comparing \pynguinstall{}, \sbst{}, \del{\codastall{}}\ins{\codastallrevised{}}, and \del{\coda{}}\ins{\codarevised{}},\ins{ from Table~\ref{tab:rq1_coverage},} we found that on average, \sbst{} achieves a branch coverage of 43.09\% within only 2.39 minutes, but just achieves 47.71\% coverage after continuing running for 17.61 minutes.
Similarly, \del{\coda{}}\ins{\codarevised{}} achieves 43.41\% branch coverage within only 3.47 minutes, but in the end just achieves 50.87\% coverage after continuing running for 16.53 minutes. \ins{From Table~\ref{tab:rq1_hard}, this trend is consistent for hard-to-cover branches as well, where \sbst{} and \del{\coda{}}\ins{\codarevised{}} quickly encounter bottlenecks in achieving further coverage improvements on hard-to-cover branches within a short time period.}
Figure~\ref{fig:coverage_trend} shows the branch-coverage trend achieved by \pynguinbase{}, \sbst{}, \codamosabase{}, and \del{\coda{}}\ins{\codarevised{}}. From this figure, \sbst{} and \del{\coda{}}\ins{\codarevised{}} can quickly cover easily reachable branches, and then encounter the coverage bottleneck to some degree, while \pynguinbase{} and \codamosabase{} continue to achieve new coverage.
This highlights the challenge posed by hard-to-cover branches and underscores the motivation behind our work.
This also confirms that the two-minute timeframe set for switching to LLMs is sufficient for \sbst{} and \del{\coda{}}\ins{\codarevised{}}, which can reflect their potential in covering hard-to-cover branches to a large extent, as extending their running time does not bring significant coverage improvement.
\ins{
Additionally, over the 20-minute period, \tech{}'s coverage continues to increase, though at a slower rate after the first 10 minutes, underscoring the importance of sufficient testing time to fully harness the testing capabilities. 
In contrast, Pynguin and \codarevised{} quickly plateau within the first few minutes, while TELPA steadily achieves higher coverage. Notably, TELPA also surpasses both Pynguin and \codarevised{} within the 10-minute time budget, demonstrating its effectiveness.
}

\begin{tcolorbox}[colback=gray!5]
\textbf{Takeaway I}: 
The state-of-the-art test generation \del{techniques}\ins{methods} (\sbst{} and \del{\coda{}}\ins{\codarevised{}}) encounter the bottleneck in coverage improvement at an early stage of testing, possibly due to those hard-to-cover branches.
\end{tcolorbox}

\ins{From Table~\ref{tab:rq1_coverage}, after}\del{After} switching to \tech{}, the branch coverage can be significantly improved according to results of \pynguinbase{} and \codamosabase{}.
Specifically, after reaching the testing time budget, \pynguinbase{} achieves an average improvement of 34.10\% over \sbst{} across all projects in terms of branch coverage, and \codamosabase{} achieves an average improvement of 25.93\% over \del{\coda{}}\ins{\codarevised{}}.
In terms of line coverage, \pynguinbase{} achieves an average improvement of 16.72\% over \sbst{} and \codamosabase{} achieves an average improvement of 9.32\% over \del{\coda{}}\ins{\codarevised{}}.
\ins{
Additionally, \pynguinbase{} outperforms \codarevised{} in terms of both branch and line coverage.
Specifically, \pynguinbase{} achieves an average branch coverage of 63.98\% while \codarevised{} achieves an average branch coverage of 50.87\%.
This also demonstrates the superiority of the prompting method and the test generation process designed in \tech{}.
Regarding hard-to-cover branches, as shown in Table~\ref{tab:rq1_hard}, both \pynguinbase{} and \codamosabase{} outperform \sbst{} and \del{\coda{}}\ins{\codarevised{}} on both scenarios, including those with complex dependencies and those with complex objects. For example, \pynguinbase{} covers 39.53\% hard-to-cover branches, outperforming \sbst{}, which covers 31.62\% branches with an improvement of 25.02\%. Similarly, \codamosabase{} covers 35.23\% hard-to-cover branches, compared to \del{\coda{}}\ins{\codarevised{}}'s 20.58\% branches, achieving an improvement of 71.18\%.
}
Furthermore, we conducted a paired Wilcoxon signed-rank test~\cite{wilcoxon} at a significance level of 0.05 between \pynguinbase{}/\codamosabase{} and \sbst{}/\del{\coda{}}\ins{\codarevised{}}, confirming \tool{}'s statistical superiority over the two state-of-the-art \del{techniques}\ins{methods} with all p-values below 0.05.

Regardless of the preceding test generation tools (\sbst{} or \del{\coda{}}\ins{\codarevised{}}), \tech{} can achieve similar effectiveness, i.e., \pynguinbase{} achieves similar branch coverage and line coverage to \codamosabase{} on average across all projects.
The results demonstrate the stable effectiveness of \tech{}.
We conducted a paired Wilcoxon signed-rank test~\cite{wilcoxon} at a significance level of 0.05 between \pynguinbase{} and \codamosabase{} in terms of both branch coverage and line coverage.
Both p-values exceed 0.05, demonstrating no statistically significant difference between \pynguinbase{} and \codamosabase{} in improving coverage.

\begin{tcolorbox}[colback=gray!5]
\textbf{Takeaway II}: 
\tech{} significantly outperforms the state-of-the-art \sbst{} and \del{\coda{}}\ins{\codarevised{}} \del{techniques} in improving both branch coverage and line coverage, regardless of the used preceding test generation tools.
\end{tcolorbox}

Compared to \sbst{}, all the LLM-based\ins{/LLM-enhanced} techniques (including \pynguinbase{}, \codamosabase{}, \del{\llmsbst{}}\ins{\llmsbstrevised{}}, \del{\llmcoda{}}\ins{\llmcodarevised{}}, \del{\coda{}}\ins{\codarevised{}}) achieve higher branch coverage and line coverage on average across all projects.
This demonstrates the effectiveness of LLMs in improving test coverage.
However, different prompting methods can largely affect the effectiveness of LLMs.
By comparing \pynguinbase{}/\codamosabase{} with \del{\llmsbst{}}\ins{\llmsbstrevised{}}/\del{\llmcoda{}}\ins{\llmcodarevised{}}, our designed prompting method specific to the challenge posed by hard-to-cover branches performs much better than the general prompting method employed for LLM-based test generation.
Specifically, the average improvement of \pynguinbase{} over \del{\llmsbst{}}\ins{\llmsbstrevised{}} is 20.19\% in terms of branch coverage and 7.61\% in terms of line coverage.
Similarly, the average improvement of \codamosabase{} over \del{\llmcoda{}}\ins{\llmcodarevised{}} is 22.02\% in terms of branch coverage and 2.22\% in terms of line coverage.
\ins{The improvement is also consistent on hard-to-cover branches. Specifically, the improvement of \pynguinbase{} over \del{\llmsbst{}}\ins{\llmsbstrevised{}} is 21.00\% and the improvement of \codamosabase{} over \del{\llmcoda{}}\ins{\llmcodarevised{}} is 21.06\% on hard-to-cover branches.}
This implies the importance of designing task-specific prompting, i.e., extracting relevant information for improving the coverage of hard-to-cover branches in \tech{}.
We conducted a paired Wilcoxon signed-rank test~\cite{wilcoxon} at a significance level of 0.05 and found that \pynguinbase{}/\codamosabase{} significantly outperforms \del{\llmsbst{}}\ins{\llmsbstrevised{}}/\del{\llmcoda{}}\ins{\llmcodarevised{}} 
by obtaining all p-values smaller than 0.05.

\begin{tcolorbox}[colback=gray!5]
\textbf{Takeaway III}: 
Incorporating LLMs into test generation helps improve test coverage compared to traditional SBST.
Designing task-specific prompting (the one in \tech{} specific to hard-to-cover branches) can further improve the effectiveness of LLM-based test generation compared to general prompting.
\end{tcolorbox}

\begin{figure}[t]   
  \centering     
  \begin{minipage}[t]{0.44\textwidth}
  \subfloat[\footnotesize \ins{Hard-to-cover branches covered by \pynguinbase{}, \llmsbstrevised{}, and \sbst{}}]{
        \includegraphics[width=\linewidth]{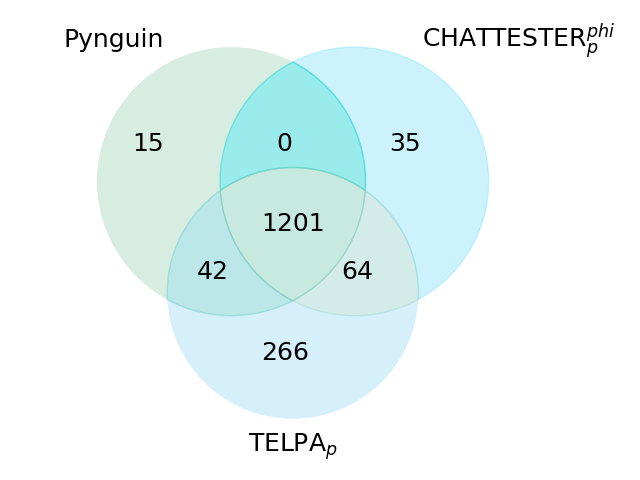}
        \label{fig:venn_pynguin}
        }
  \end{minipage}
  \hspace{0.2cm}
  \begin{minipage}[t]{0.44\textwidth}
  \subfloat[\footnotesize \ins{Hard-to-cover branches covered by \codamosabase{}, \llmcodarevised{}, and \codarevised{}}]{
      \includegraphics[width=\linewidth]{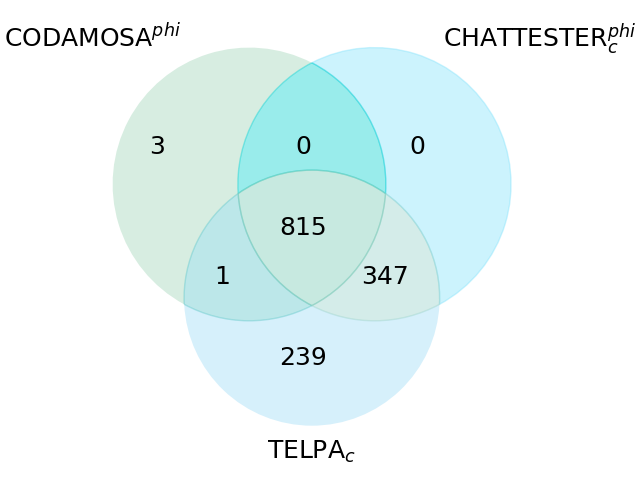}
        \label{fig:venn_codamosa}
    }
  \end{minipage}
  \\
    \begin{minipage}[t]{0.44\textwidth}
  \subfloat[\footnotesize \ins{Easy-to-cover branches covered by \pynguinbase{}, \llmsbstrevised{}, and \sbst{}}]{
        \includegraphics[width=\linewidth]{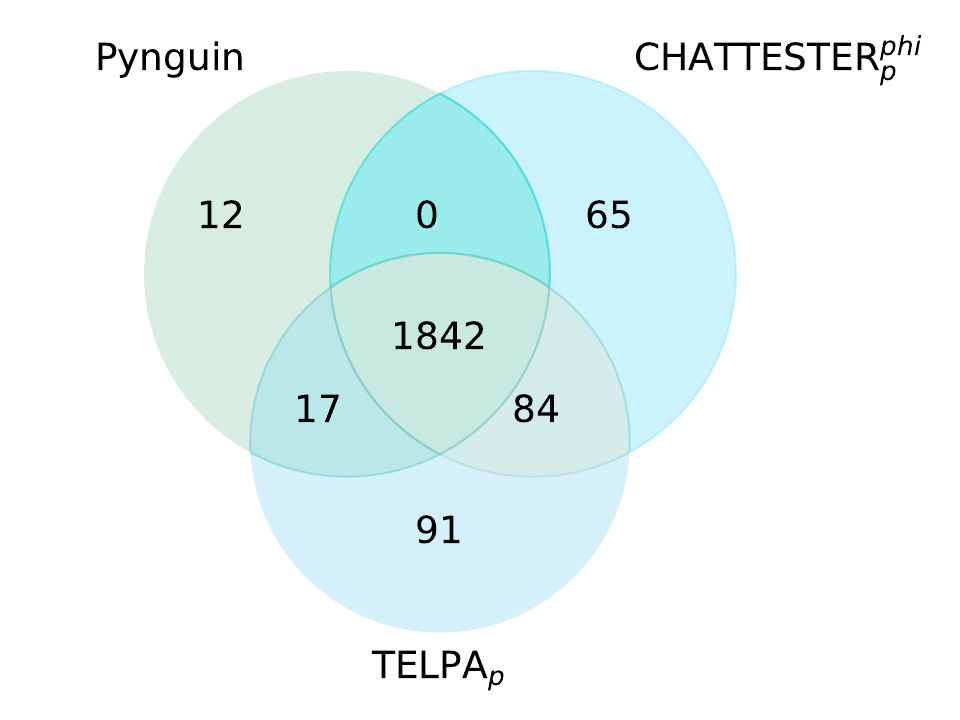}
        \label{fig:venn_pynguin_easy}
        }
  \end{minipage}
  \hspace{0.2cm}
  \begin{minipage}[t]{0.44\textwidth}
  \subfloat[\footnotesize \ins{Easy-to-cover branches covered by \codamosabase{}, \llmcodarevised{}, and \codarevised{}}]{
      \includegraphics[width=\linewidth]{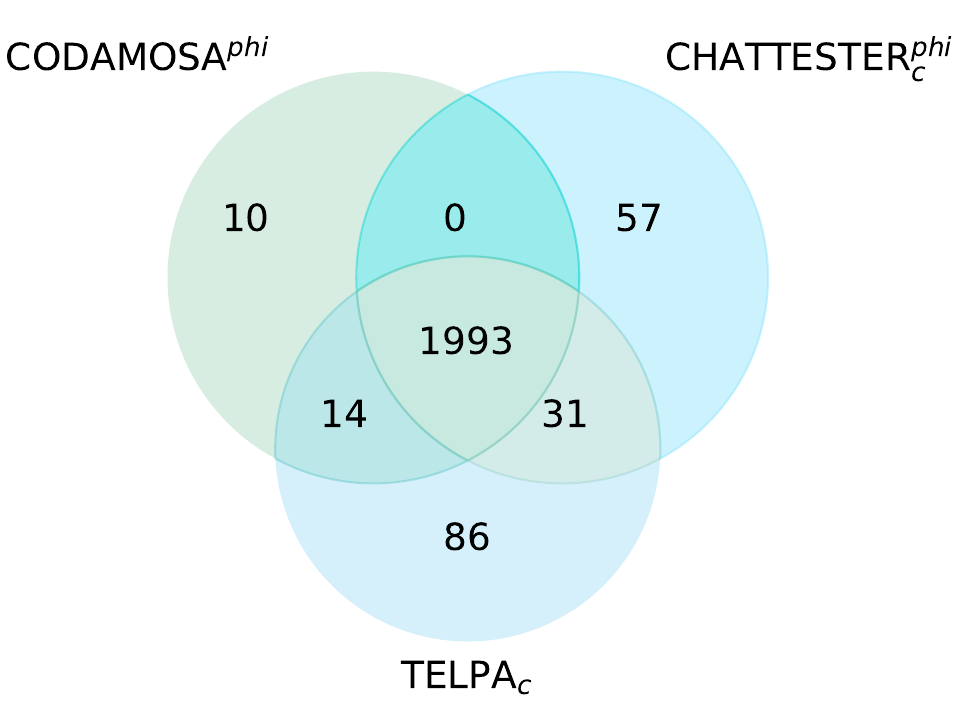}
        \label{fig:venn_codamosa_easy}
    }
  \end{minipage}
  \caption{\ins{Different branches covered by different techniques}}
\label{fig:venn_branch}
\end{figure}

\ins{
 To gain deeper insights into the specific branches covered by each method, we analyzed the overlap in covered branches across different methods, as visualized in Figure~\ref{fig:venn_branch}. The figure demonstrates that all studied techniques perform well on easy-to-cover branches. However, \tech{} not only covers nearly all the hard-to-cover branches covered by other methods but also covers a significant number of unique hard-to-cover branches. This underscores \tech{}'s contribution to enhancing coverage of hard-to-cover branches and complementing existing methods.
}

\begin{tcolorbox}[colback=gray!5]
\ins{
\textbf{Takeaway IV}: \tech{} excels in covering hard-to-cover branches, including those missed by other methods, which highlights its potential to complement existing test generation techniques.
}
\end{tcolorbox}

\subsection{RQ2: Contribution of Each Component in \tech{}}

\begin{table}[t]
\setlength{\tabcolsep}{2.5pt}
\caption{\ins{Comparison between \tech{} and its variants in terms of branch and line coverage on average across all the projects}}
\centering
\small
\label{tab:rq2_abalation}
\renewcommand{\arraystretch}{1.5}
\resizebox{0.75\linewidth}{!}
{
\begin{threeparttable}
\ins{
\begin{tabular}{l|rrrrrrrrr}
\toprule
\multicolumn{1}{l|}{\textbf{Cov.}} & 
\multicolumn{1}{c}{\textbf{\nocallsbst{}}} & 
\multicolumn{1}{c}{\textbf{\nobranchsbst{}}} & 
\multicolumn{1}{c}{\textbf{\noshotsbst{}}} & 
\multicolumn{1}{c}{\textbf{\nofeedbacksbst{}}} & 
\multicolumn{1}{c}{\textbf{\randomshotsbst{}}} & 
\multicolumn{1}{c}{\textbf{\nofilteringsbst{}}} & 
\multicolumn{1}{c}{\textbf{\nocotsbst{}}} &
\multicolumn{1}{c}{\textbf{\pynguinbase{}}} 
\\
\midrule
Branch Cov. & 55.87\% & 58.74\% & 55.17\% & 58.34\% & 58.69\% & 52.80\% & 59.84\%  & \textbf{63.98\%}  \\
Line Cov.   & 70.67\% & 71.80\% & 70.46\% & 71.65\% & 71.07\% & 69.17\% & 72.29\%  & \textbf{75.54\%}  \\
\midrule
\multicolumn{1}{l|}{\textbf{Cov.}} &
\multicolumn{1}{c}{\textbf{\nocallcoda{}}} & 
\multicolumn{1}{c}{\textbf{\nobranchcoda{}}} & 
\multicolumn{1}{c}{\textbf{\noshotcoda{}}} & 
\multicolumn{1}{c}{\textbf{\nofeedbackcoda{}}} & 
\multicolumn{1}{c}{\textbf{\randomshotcoda{}}} & 
\multicolumn{1}{c}{\textbf{\nofilteringcoda{}}} & 
\multicolumn{1}{c}{\textbf{\nocotcoda{}}} &
\multicolumn{1}{c}{\textbf{\codamosabase{}}} \\
\midrule
Branch Cov. &  56.68\% & 60.45\% & 56.12\% & 57.77\% & 57.32\% & 48.89\% & 60.12\%  & \textbf{64.06\%} \\
Line Cov.   & 72.52\% & 73.78\% & 72.24\% & 72.30\% & 71.89\% & 66.88\% & 72.48\%          & \textbf{74.25\%} \\
\bottomrule
\end{tabular}
}
\end{threeparttable}
}
\end{table}

To investigate the contribution of each component in \tool{}, we compared \tool{} with a set of its variants (introduced in Section~\ref{sec:rq2_baselines}). 
Table~\ref{tab:rq2_abalation} presents the comparison results in terms of average branch coverage and average line coverage across all the projects.
As can be seen from Table~\ref{tab:rq2_abalation}, both \pynguinbase{} and \codamosabase{} perform better than their corresponding variants.
We further performed a paired Wilcoxon signed-rank test at the significance level of 0.05 to investigate whether \pynguinbase{}/\codamosabase{} significantly outperforms each variant across all projects in terms of branch coverage and line coverage.
All p-values are smaller than 0.05, demonstrating the statistically significant contribution of each component in \tech{},
regardless of the used preceding test generation tools.

\nofiltering{} performs the worst among all variants. 
This is because the unfiltered invocation sequences can be excessively long and complex, posing significant challenges for LLMs to comprehend the complex sequence, which in turn reduces the effectiveness of test generation. The result demonstrates the contribution of sequence filtering, which could also balance the sequence content and computational burden well.
In particular, except \nofiltering{}, all other variants still outperform \llm{} even when some component is removed from \tech{}.
This demonstrates the effectiveness of using part of information specific to the challenges posed by hard-to-cover branches, compared to directly integrating 
unnecessary or insufficient context for prompting, which also reflects the negative influence of introducing irrelevant information on LLMs.

We also investigated whether our coverage-based counter-example sampling approach %
contributes to the overall effectiveness of \tech{} by comparing with \randomshot{} that randomly selects counter-examples.
Table~\ref{tab:rq2_abalation} shows that \randomshot{} outperforms \noshot{} regardless of the preceding test generation tools.
This result further confirms the importance of using counter-examples even if they are just selected randomly.
Moreover, \tech{} significantly outperforms \randomshot{} regardless of the preceding test generation tools.
For example, \pynguinbase{} improves \randomshotsbst{} by 9.01\% in terms of average branch coverage achieved, while \codamosabase{} improves \randomshotcoda{} by 11.76\%.
The results demonstrate the importance of a diverse set of counter-examples, which can be well achieved by our coverage-based sampling strategy.
\ins{
Additionally, TELPA$_{\textit{ncot}}$ reduced branch coverage by 6.3\% on average. The results empirically demonstrate that summarizing the functionality beforehand enhances test generation effectiveness.
}

\begin{tcolorbox}[colback=gray!5]
\textbf{Takeaway \del{IV}\ins{V}}: 
Each component in \tech{} contributes to the overall effectiveness significantly, regardless of the preceding test generation tools used by \tech{}.
\end{tcolorbox}

\vspace{2mm}

\subsection{RQ3: \tech's Effectiveness with Different Configurations}
\label{sec:rq3_configuration}
\begin{table}[t]
\setlength{\tabcolsep}{2.5pt}
\caption{Effectiveness of \tech{} under different configurations in terms of average branch and line coverage}
\centering
\label{tab:rq3_scratch}
\renewcommand{\arraystretch}{1.4}
\resizebox{0.75\linewidth}{!}
{
\begin{threeparttable}
\begin{tabular}{l|rrr|rr|rr}
\toprule
\textbf{Cov.} &
\textbf{\sbst{}} &
\textbf{\coda{}} &
\textbf{\techzero{}} &
\textbf{\deepseeksbst{}} &
\textbf{\pynguinbase{}} &
\textbf{\deepseekcoda{}} &
\textbf{\codamosabase{}}
\\
\midrule
Branch Cov. &  47.71\% & 50.87\% & \textbf{52.75\%} &  60.34\% & \textbf{63.98\%} & 59.79\% & \textbf{64.06\%} \\
Line Cov.  & 64.72\% &  67.92\%  & \textbf{68.26\%} & 73.96\% & \textbf{75.54\%} & 74.08\%  & \textbf{74.25\%} \\
\bottomrule
\end{tabular}
\end{threeparttable}
}
\end{table}

\ins{\tech{} uses existing tools for initial test generation, and utilizes an LLM to generate new tests. In this RQ, we examine how the adoption of existing tools and the choice of the LLM can impact the performance.}
We first investigated the effectiveness of \techzero{} that removes the preceding test generation tool and activates \tech{} from scratch.
By comparing \techzero{} with \del{\coda{}}\ins{\codarevised{}} and \sbst{} in Table~\ref{tab:rq3_scratch}, we can see that the \techzero{} can improve \del{\coda{}}\ins{\codarevised{}} and \sbst{} by 3.70\% and 10.56\% in terms of average branch coverage and 0.50\% and 5.47\% in terms of average line coverage.
The improvements are not huge, especially compared to the differences between \codamosabase{} and \del{\coda{}}\ins{\codarevised{}} or between \pynguinbase{} and \sbst{}.
This is expected, as the time needed for LLMs to generate each test cases is normally much more than that required by SBST techniques. \techzero{} leverages LLMs for even easy-to-cover branches, preventing it from unleashing its full potential to reach hard-to-cover branches.
Instead, \del{\coda{}}\ins{\codarevised{}} alternately invokes Pynguin and LLMs for test generation, reducing the time needed by LLMs to a large extent, while \sbst{} is also efficient as it does not rely on LLMs.
Hence, within the same limited testing time, \techzero{} does not achieve significant coverage improvement over \del{\coda{}}\ins{\codarevised{}} and \sbst{}.
This justifies our design choice of adopting preceding test generation tools to balance effectiveness and efficiency.

By examining the data, we found that 603 methods were not covered by \del{\coda{}}\ins{\codarevised{}} and \sbst{} at all within the testing time budget. For these methods, \techzero{} instead can effectively improve the coverage for 259 of them. 
On average across these methods, \codamosabase{} achieves 21.86\% branch coverage and 39.99\% line coverage, while \pynguinbase{} achieves 17.14\% branch coverage and 43.80\% line coverage. 
The results further confirm the ability of \tech{} in improving the coverage of hard-to-coverage branches.

\begin{tcolorbox}[colback=gray!5]
\textbf{Takeaway \del{V}\ins{VI}}: 
Applying \tech{} to easy-to-cover branches is not recommended due to the non-negligible time required by LLMs. Our designed usage scenario for \tech{} balances effectiveness and efficiency well.
\end{tcolorbox}

\vspace{2mm}

We then investigated the effectiveness of \deepseeksbst{} and \deepseekcoda{} that employ \textit{DeepSeek-Coder-6.7B-instruct}~\cite{deepseek} instead of \textit{Phind-CodeLlama-34B-v2}~\cite{phind}.
From Table~\ref{tab:rq3_scratch},
we found although \deepseeksbst{}/\deepseekcoda{} exhibits small performance reduction compared to \pynguinbase{}/\codamosabase{}, the former still largely outperforms \sbst{}/\del{\coda{}}\ins{\codarevised{}}. 
For example, \deepseeksbst{}/\deepseekcoda{} improves \sbst{}/\del{\coda{}}\ins{\codarevised{}} by 26.47\% and 17.53\% in terms of average branch coverage.
To investigate the effectiveness of \tech{} when using more advanced LLMs, we evaluated \tech{} on GPT-4. 
We randomly sampled 100 target methods from the benchmark used in previous RQs for evaluation due to the high cost of invoking GPT-4's APIs.
From Table~\ref{tab:rq3_gpt}, \gptsbst{} exhibits similar effectiveness to \pynguinbase{} in terms of both average branch coverage and average line coverage on the same 100 sampled methods. 
This is attributed to the strength of Phind-CodeLlama-34B-v2, a powerful LLM that is specifically trained on code corpus and exhibits close performance to GPT-4 on many code-related tasks~\cite{huggingface_leaderboard}.
The results demonstrate the generalizability of \tech{} under different LLMs to some degree. 

\begin{table}[t]
\setlength{\tabcolsep}{2.5pt}
\caption{Effectiveness of \tech{} using GPT-4 evaluated on 100 randomly sampled target methods}
\centering
\label{tab:rq3_gpt}
\renewcommand{\arraystretch}{1.5}
\resizebox{0.6\linewidth}{!}
{
\begin{threeparttable}
\begin{tabular}{l|rr|rrr}
\toprule
\textbf{Cov.} &
\textbf{\sbst{}} &
\textbf{\coda{}} &
\textbf{\deepseeksbst{}} &
\textbf{\pynguinbase{}} &
\textbf{\gptsbst{}}
\\
\midrule
Branch Cov. &  26.86\% & 27.93\% & 35.68\% &  37.76\% & 38.43\% \\
Line Cov.  & 42.13\% &  45.75\%  & 53.66\% & 55.97\% & 56.08\% \\
\bottomrule
\end{tabular}
\end{threeparttable}
}
\end{table}

\begin{tcolorbox}[colback=gray!5]
\textbf{Takeaway \del{VI}\ins{VII}}: 
Leveraging small-scale LLMs in \tech{} can also largely improve test coverage, demonstrating the generalizability of \tech{}.
By incorporating more advanced LLMs, the effectiveness of \tech{} is further improved.
\end{tcolorbox}

\vspace{2mm}

\subsection{\ins{RQ4: Correctness of Unit Tests Generated by \tech{}}}
\label{sec:rq4_correctness}
\begin{table*}[t]
\setlength{\tabcolsep}{2.5pt}
\caption{\ins{Comparison among \tech{}, \llmrevised{}, \sbst{} and \codarevised{} in terms of correctness}}
\centering
\small
\label{tab:rq4}
\renewcommand{\arraystretch}{1.4}
\resizebox{0.99\linewidth}{!}
{
\begin{threeparttable}
\ins{
\begin{tabular}{l|rrr|rrr}
\toprule
\textbf{Metric}  & \textbf{\sbst{}} & \textbf{\llmsbstrevised{}} & \textbf{\pynguinbase{}} & \textbf{\codarevised{}} & \textbf{\llmcodarevised{}} & \textbf{\codamosabase{}} \\
\midrule
Execution Pass Rate &  97.44\% &  78.57\% & 62.48\% &  97.66\% & 78.82\% & 62.07\% \\
\bottomrule
\end{tabular}
}
\end{threeparttable}
}
\end{table*}

\ins{First, we found that all techniques achieve 100\% syntax correctness, demonstrating their ability to generate syntactically valid tests without errors. Then, Table~\ref{tab:rq4} compares the execution pass rate of unit tests generated by \sbst{}, \codarevised{}, \llmrevised{} and \tool{}.
From the table, both the LLM-based \tool{} and \llmrevised{} exhibit lower execution pass rates than the search-based \sbst{}, indicating the challenge of LLMs in generating semantically valid and executable tests.
Note that although \codarevised{} also relies on LLMs, it only leverages LLMs to generate seed tests for \sbst{}, and most of the tests are ultimately generated by \sbst{}, which results in comparable pass rates between \codarevised{} and \sbst{}.
\tool{} exhibit lower execution pass rates than \llmrevised{}. This is because \tool{} specifically targets hard-to-cover branches, which often require constructing complex objects and resolving intricate inter-procedural dependencies. These tests are inherently more challenging to execute successfully due to their complexity, involving scenarios prone to runtime errors like incorrect object construction or missing dependencies. In contrast, \llmrevised{} generates simpler tests that are less effective at covering complex branches. 
For example, \llmrevised{}'s tests often cover straightforward branches with minimal object construction and fewer inter-procedural dependencies, resulting in higher pass rates but lower overall branch coverage. 
Specifically, \pynguinbase{} and \codamosabase{} cover 21.00\% and 21.06\% more hard-to-cover branches than \llmsbstrevised{} and \llmcodarevised{}, respectively.
}

\ins{
We further analyzed the distribution of errors in the generated tests. For TELPA, the tests predominantly fail due to AssertionError (32.0\%) and AttributeError (33.5\%). 
AssertionError occurs when the test fails an assertion, which is common in tests targeting complex branches where the expected behavior is harder to predict or verify.
AttributeError often happens when the test attempts to access an attribute or method on an object that is not properly constructed or initialized, which is a common issue when dealing with complex object dependencies.
Overall, these results confirm that while \tool{} generates more complex tests that may have lower execution pass rates, these tests are often essential for achieving higher branch coverage, particularly for hard-to-cover branches involving complex objects and inter-procedural dependencies.
}

\begin{tcolorbox}[colback=gray!5]
\ins{
\textbf{Takeaway VIII}: \tech{} generates more complex unit tests to target hard-to-cover branches, which, while resulting in slightly lower execution pass rates, enables it to achieve significantly higher branch coverage.
}
\end{tcolorbox}

\subsection{RQ5: Generalizability of \tool{}}
\label{sec:java_evaluation}

\begin{table*}[t]
\setlength{\tabcolsep}{2.5pt}
\caption{\ins{Comparison between \tech{} and \evo{} in terms of branch coverage on all branches}}
\centering
\small
\label{tab:evo_coverage}
\renewcommand{\arraystretch}{1.4}
\resizebox{0.4\linewidth}{!}
{
\begin{threeparttable}
\ins{
\begin{tabular}{l|rr}
\toprule
\textbf{Project}  & \textbf{\evo{}} & \textbf{\evosuitebase{}} \\
\midrule
Chart               & 48.64\%  & 52.45\%  \\
Lang                & 85.00\%  & 88.71\%  \\
Math                & 73.57\%  & 77.51\%  \\
Time                & 43.49\%  & 55.40\%  \\
\midrule
Average Branch Cov. & 62.68\%  & 68.52\%  \\
Average Line Cov.   & 73.09\%  & 76.85\%  \\
\bottomrule
\end{tabular}
}
\end{threeparttable}
}
\end{table*}

\begin{table*}[t]
\setlength{\tabcolsep}{2.5pt}
\caption{\ins{Comparison between \tech{} and \evo{} in terms of branch coverage on hard-to-cover branches}}
\centering
\small
\label{tab:evo_hard}
\renewcommand{\arraystretch}{1.4}
\resizebox{0.5\linewidth}{!}
{
\begin{threeparttable}
\ins{
\begin{tabular}{c|rr}
\toprule
\textbf{Branch Type}  & \textbf{\evo{}} & \textbf{\evosuitebase{}} \\
\midrule
Branches with complex dependencies & 66.53\% & 77.42\% \\
Branches with complex objects       & 56.79\% & 74.90\% \\
\midrule
\makecell{All hard-to-cover branches \\ (union of the above two types)}    & 57.55\% & 74.59\% \\
\bottomrule
\end{tabular}
}
\end{threeparttable}
}
\end{table*}

\insminor{
}

\insminor{
Table~\ref{tab:evo_coverage} presents the comparison results for \evosuitebase{} vs. \evo{} on each project in terms of branch coverage on all branches. 
The last two rows present the average branch coverage and line coverage across all projects.
Table~\ref{tab:evo_hard} presents the comparison results on hard-to-cover branches.
From Table~\ref{tab:evo_coverage} and Table~\ref{tab:evo_hard}, \evosuitebase{} achieves consistently higher branch coverage than \evo{}.
Specifically, on average, TELPA improves \evo{} by 9.32\% on all branches and 29.60\% on hard-to-cover branches in terms of branch coverage.
Note that the improvement on all branches is smaller than the improvement TELPA achieves over Pynguin}\delminor{, this is because \evo{} incorporates more advanced static analysis methods, which allow \evo{} to perform better than Pynguin in many cases}. \insminor{
One possible explanation is that \evo{} already incorporates more advanced static analysis techniques than Pynguin, potentially diminishing the additional benefits offered by \tech{}. Moreover, other factors, such as differences in programming languages, testing strategies, and even tool implementation details, may also contribute to this discrepancy.}
\insminor{Even though, they are still limited in handling intricate inter-procedural dependencies and dynamically computed values, which is confirmed by the larger improvement achieved by \tech{} on hard-to-cover branches. 
The constant pool, for instance, may struggle when required values are not explicitly defined as constants but are derived from complex object construction logic. 
Similarly, testability transformations excels in cases where flags create coarse fitness landscapes in search-based approaches, making it difficult
to find paths leading to high-fitness regions.
However, when the program contains complex inter-dependencies or complex object constructions, testability transformation alone may not provide sufficient guidance. 
The transformation does not inherently handle the semantics of such complex dependencies, which are crucial for generating tests in challenging branches. 
The results reflects the fact that \evo{} is already a highly-optimized tool, but \tech{} still provides meaningful improvements in terms of code
coverage, particularly for those much more challenging branches.
}

\begin{tcolorbox}[colback=gray!5]
\insminor{
\textbf{Takeaway IX}: Even when compared to a highly optimized tool (i.e., \evo{}) on a statically typed language (i.e., Java), \tech{} consistently improves coverage, particularly on hard-to-cover branches, demonstrating its generalizability across different languages.
}
\end{tcolorbox}

\subsection{Threats to Validity}
\label{sec:threats}

The threats mainly lie in parameter settings in \tech{} \del{and metrics used}\ins{, metrics and benchmarks used, and the randomness}.
By default, we set the timeframe to two minutes for switching from preceding test generation tools to our LLM-based test generation, the LLM temperature to \del{1}\ins{0}, and the testing time budget to 20 minutes on each software module.
Both the timeframe and the testing time budget are longer than the settings in the existing studies~\cite{pynguin,codamosa}, demonstrating higher sufficiency of reaching coverage bottlenecks and evaluating test effectiveness.
The results shown in Figure~\ref{fig:coverage_trend} also confirm it.
\del{Higher}\ins{Lower} temperature allows \tech{} to receive more deterministic responses from LLMs, which is commonly employed in various tasks~\cite{chen2021evaluating,coco,tian2025fixing, yang2024dependency}. 
In the future, we will investigate the influence of different settings on the effectiveness of \tech{}.
Regarding metrics, we have employed the widely-used branch coverage and line coverage, which are also aligned with our goal (i.e., reaching hard-to-cover branches).
We also analyzed the overhead of our static program analysis: 
on average, the time spent on each module is 1.5 seconds. 
This overhead is acceptable given its effectiveness. Moreover, this process is performed offline and it could be further accelerated via parallel executions. 

Following the existing work~\cite{codamosa}, we did not investigate the bug detection effectiveness because test cases must include proper assertions to reveal bugs. 
However, \tech{} focuses on improving branch coverage rather than generating assertions, and various assertion generation approaches~\cite{toga, assertion_1} can be applied to the test cases generated by \tech{}.
In the future, we will use more metrics to evaluate generated tests more sufficiently.

\ins{Regarding benchmarks, one potential threat to the validity of our results is the dynamically typed nature of Python, which presents unique challenges for automated test generation. In dynamically typed languages, type information is not explicitly available at compile time.
This makes it difficult for automated test generation tools to infer correct types for object construction and method invocations, potentially leading to ineffective or invalid tests. 
Furthermore, automated test generation tools for Python, such as Pynguin and CODAMOSA, lack the advanced static analysis techniques used in more mature tools like EvoSuite.
This may lead to ineffective or invalid tests and bias the evaluation results. To mitigate this threat, we adapted \tool{} to statically typed languages (i.e., Java) and evaluate its effectiveness against \evo{} on a widely-used Java benchmark Defects4J~\cite{d4j} in \delminor{Section~\ref{sec:discussion}}\insminor{Section~\ref{sec:java_evaluation}}. The results demonstrate that \tool{} consistently outperforms \evo{} on Java projects, which helps mitigating this threat to some extent.} 

\insminor{
\begin{table*}[t]
\setlength{\tabcolsep}{2.5pt}
\caption{\ins{Comparison between \llmbase{} and \llmrevised{} in terms of branch coverage on all branches}}
\centering
\small
\label{tab:data_leakage}
\renewcommand{\arraystretch}{1.4}
\resizebox{0.4\linewidth}{!}
{
\begin{threeparttable}
\ins{
\begin{tabular}{l|rr}
\toprule
\textbf{Project}  & \textbf{\llmrevised{}} & \textbf{\llmbase{}} \\
\midrule
MicroService        & 26.47\%    & 64.71\% \\
PATool              & 89.08\%    & 90.78\% \\
DAService           & 42.71\%    & 53.65\% \\
\midrule
Average Branch Cov. & 52.75\%    & 69.71\% \\
Average Line Cov.   & 59.43\%    & 72.71\% \\
\bottomrule
\end{tabular}
}
\end{threeparttable}
}
\end{table*}

}
\insminor{
Furthermore, the projects used for evaluation may appear in the training data of the LLMs used in our study, which can cause the data leakage problem and bias the results~\cite{data_leak}. 
Particularly, following the guidelines provided by previous work~\cite{data_leak}, we also adopted three internal Java projects provided by our industrial partner (Huawei, a global leader company in IT), which can largely reduce the threat from the potential data leakage of open-source projects.
The three industrial projects have different functionalities, i.e., a program analysis toolkit, an online micro-service system, and a data analysis framework involving parallel computing and adaptation of design patterns. For ease of presentation, we refer to the three projects as PATool, Microservice and DAService in the following sections. Due to the company policy, we are unable to disclose further details.
We constructed \llmbase{}, which leverages \llmrevised{} for preceding test generation and
compared it with \llmrevised{}. We did not leverage \evo{} since the three industrial projects use Java 17, but \evo{} supports up to Java 11.
}

\insminor{
Table~\ref{tab:data_leakage} shows the comparison results for \llmbase{} vs. \llmrevised{} on each project in terms of branch coverage.
From the table, \llmbase{} achieves significantly higher branch coverage than \llmrevised{} across all three industrial projects. On average, \llmbase{} improves the branch coverage from 52.75\% to 69.71\% and the line coverage from 59.43\% to 72.71\%. 
The improvements of \tech{} on these unseen projects highlight its generalizability and effectiveness in real-world scenarios, independent of potential data leakage from open-source training data.
}

\ins{
Regarding randomness, following the previous guideline~\cite{repeat}, we repeated all the experiments for 20 times\insminor{, including experiments on Java and those addressing data leakage threats}. 
With 20 runs, \tech{} exhibited a standard deviation of 1.8\%,\insminor{  demonstrating consistency. Additionally, we conducted a statistical significance test using the Kruskal–Wallis H test~\cite{kwhtest}. The result showed no significant differences across runs (\textit{p} $\textgreater$ 0.05),} indicating stable performance and mitigating this threat to some extent.
}

\section{Case Study}

\ins{
\delminor{To illustrate the scenarios where \tool{} targets and can cover}\insminor{To illustrate the scenarios that \tool{} specifically targets and is capable of covering}, we present two cases that highlight its ability to address inter-procedural dependencies and complex object construction.
}

\begin{figure}[t]
  \centering
  \includegraphics[width=0.85\linewidth]{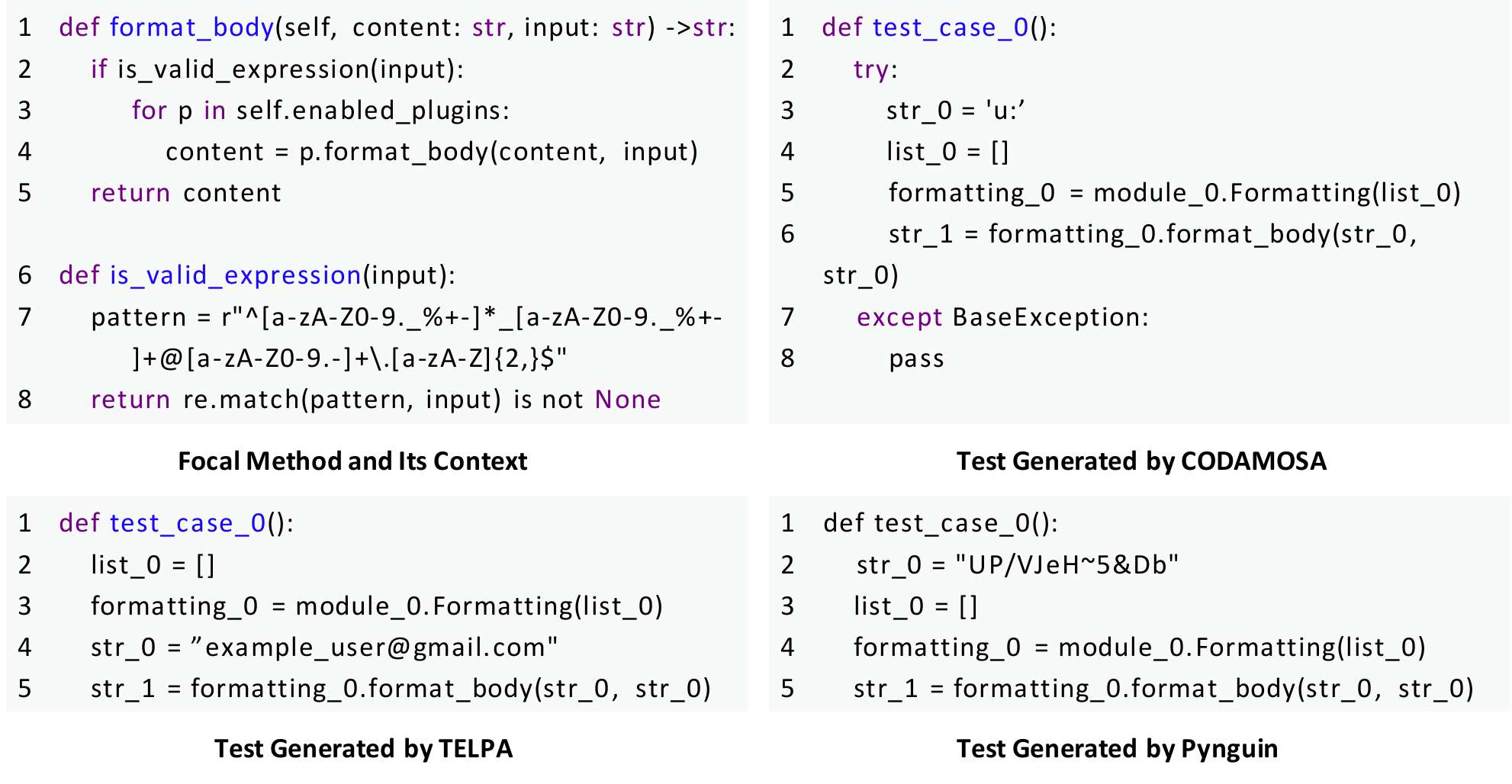}
  \caption{\ins{Case for complex dependencies}}
  \label{fig:case_study_1}
\end{figure}

\ins{
The first case is shown in Figure~\ref{fig:case_study_1}, which demonstrates how \tech{} addresses the challenge posed by complex dependencies using branch dependency analysis. The target method \texttt{format\_body} processes input content using enabled plugins only if the input matches a specific date pattern validated by \texttt{is\_valid\_expression}. 
To cover the branch where plugins are applied, the generated test must provide an input that satisfies \texttt{is\_valid\_expression}, which requires understanding the regex pattern.
Baseline tools, such as \sbst{}, generate random inputs (e.g., "UP/VJeH~5\&Db") that fail the regex check, leaving the target branch uncovered. In contrast, \tech{} leverages branch dependency analysis to identify the dependency between \texttt{format\_body} and \texttt{is\_valid\_expression}. By including the source code of \texttt{is\_valid\_expression} in the prompt, \tech{} guides the LLM to infer that valid inputs must adhere to the regex pattern, which requires an underscore in the account part of an email address (e.g., ``example\_user@gmail.com'').
This allows \tech{} to generate tests that invoke the plugin branch, while baseline tools fail due to their inability to infer valid inputs. Notably, the test generated by \llm{} is omitted here, as it differs from the \sbst{}-generated test only in a single random value that still fails the regex check.
}

\begin{figure}[t]
  \centering
  \includegraphics[width=0.85\linewidth]{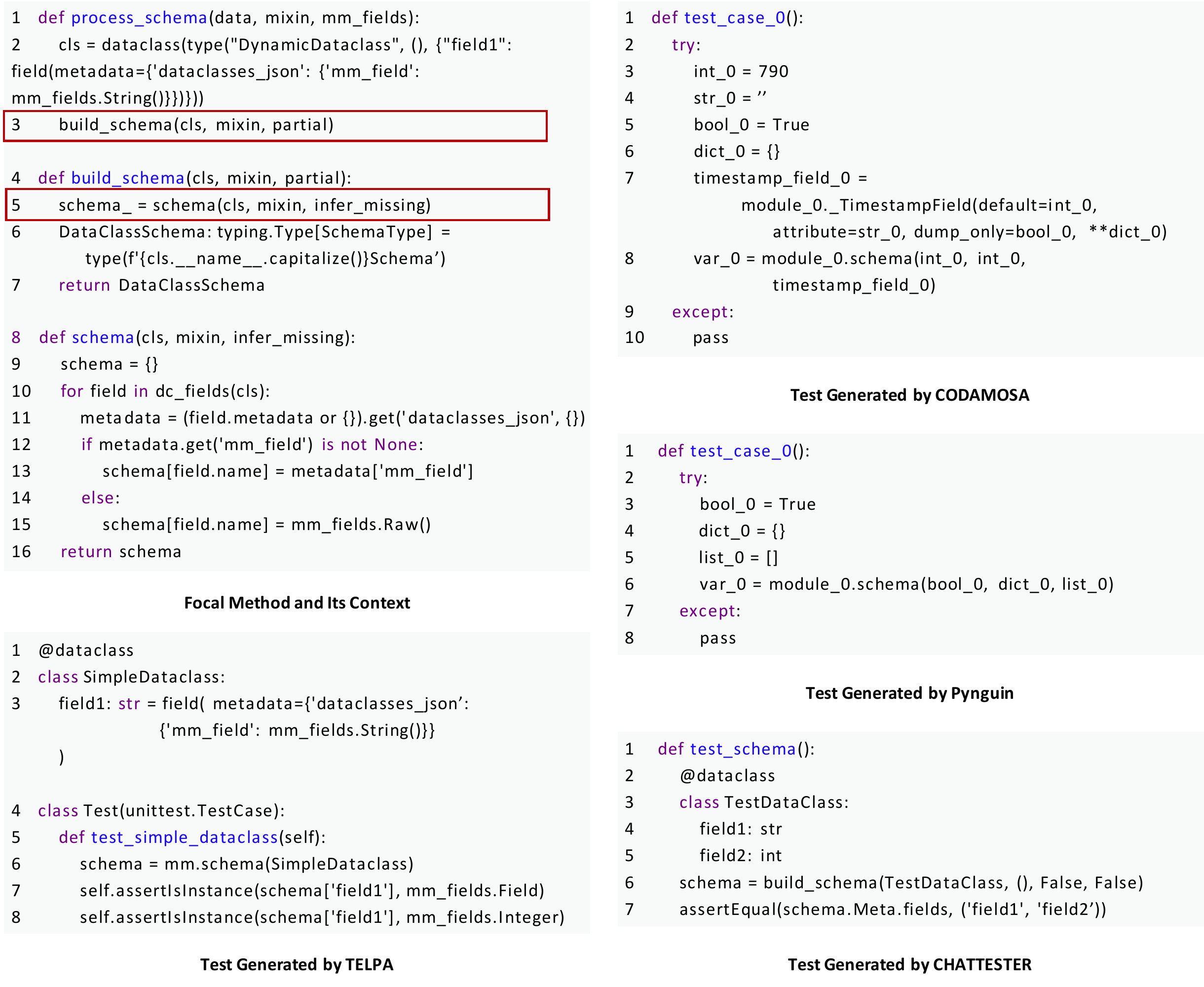}
  \caption{\ins{Case for complex object construction}}
  \label{fig:case_study_2}
\end{figure}

\ins{
The second case is shown in Figure~\ref{fig:case_study_2}, which demonstrates how TELPA addresses the challenge
posed by complex object construction using object construction analysis.
In this case, the focal method is the \texttt{schema} function, which processes a dataclass (\texttt{cls}) and constructs a schema based on the metadata of its fields. Specifically, it checks if the \texttt{metadata} dictionary of each field contains a key \texttt{dataclasses\_json} and uses the associated \texttt{mm\_field} to populate the schema. Constructing a valid \texttt{cls} object with the correct metadata structure is critical for covering the branch where \texttt{metadata.get(mm\_field)} is not \texttt{None}. Baseline tools struggle with this task: \sbst{} and \coda{} generate invalid inputs (e.g., integers, booleans) that do not match the expected \texttt{cls} type, while \llm{} constructs a valid dataclass but omits the required \texttt{dataclasses\_json} metadata. \tool{} addresses this challenge by performing object construction analysis to trace the call chain \texttt{process\_schema} $\rightarrow$ \texttt{build\_schema} $\rightarrow$ \texttt{schema}. It extracts the logic from \texttt{process\_schema}, where \texttt{cls} is dynamically instantiated with the required metadata, and incorporates this context into the prompt. As a result, \tech{} generates a valid test case with the required metadata structure, ensuring coverage of the target branch.

These cases demonstrate how \tech{} leverages program analysis to address challenges proposed by complex dependencies and object constructions. By understanding inter-procedural dependencies and the complex object construction logic, \tech{} generates tests that can cover those challenging hard-to-cover branches.
}

\section{Discussion}
\label{sec:discussion}

\delminor{
\textbf{Generalizability.}
From our study, the generalizability of \tech{} has been demonstrated to a certain extent by evaluating the performance on 27 projects and considering two types of preceding test generation tools (\sbst{} and \coda{}).
}

\delminor{
Particularly, we further evaluated \tech{} on four real-world open-source projects using the JUnit framework~\cite{junit} in the widely-used Defects4J benchmark~\cite{d4j} due to the significant popularity of Java and JUnit  (i.e., Chart, Time, Lang, and Math). 
Unlike Python, Java is a statically typed language. Automated test case generation for dynamically typed languages often struggles with object creation, as obtaining detailed type information can be challenging. In contrast, Java’s static typing allows for more precise and efficient handling of type information, making it an ideal choice for evaluating the effectiveness of TELPA on static type systems.
Here, we excluded Closure since it does not use the JUnit framework and involves significant testing costs.
In total, the benchmark contains 9,391 branches, of which 5,494 are classified as
hard-to-cover. Specifically, 2,453 branches involve complex dependencies, while 4,842 branches are associated with complex objects.
We constructed \evosuitebase{}, which leverages the state-of-the-art search-based test generation tool for Java (i.e., \evo{}~\cite{evo}) for preceding test generation and compared it with \evo{}. 
As one of the most prominent search-based techniques, \evo{} employs evolutionary algorithms to generate unit tests aimed at maximizing code coverage. Additionally, it integrates advanced program analysis techniques to further enhance its effectiveness, including leveraging a constant pool for generating test inputs and applying testability transformations to improve guidance during the test generation process, etc.
All the experimental settings are the same with RQ1.
}

\delminor{
Table~\ref{tab:evo_coverage} presents the comparison results for \evosuitebase{} vs. \evo{} on each project in terms of branch coverage on all branches. 
The last two rows present the average branch coverage and line coverage across all projects.
Table~\ref{tab:evo_hard} presents the comparison results on hard-to-cover branches.
From Table~\ref{tab:evo_coverage} and Table~\ref{tab:evo_hard}, \evosuitebase{} achieves consistently higher branch coverage than \evo{}.
Specifically, on average, TELPA improves \evo{} by 9.32\% on all branches and 29.60\% on hard-to-cover branches in terms of branch coverage.
Note that the improvement on all branches is smaller than the improvement TELPA achieves over Pynguin, this is because \evo{} incorporates more advanced static analysis methods, which allow \evo{} to perform better than Pynguin in many cases. 
However, they are still limited in handling intricate inter-procedural dependencies and dynamically computed values, which is confirmed by the larger improvement achieved by \tech{} on hard-to-cover branches. 
The constant pool, for instance, may struggle when required values are not explicitly defined as constants but are derived from complex object construction logic. 
Similarly, testability transformations excels in cases where flags create coarse fitness landscapes in search-based approaches, making it difficult
to find paths leading to high-fitness regions.
However, when the program contains complex inter-dependencies or complex object constructions, testability transformation alone may not provide sufficient guidance. 
The transformation does not inherently handle the semantics of such complex dependencies, which are crucial for generating tests in challenging branches. 
The results reflects the fact that \evo{} is already a highly-optimized tool, but \tech{} still provides meaningful improvements in terms of code
coverage, particularly for those much more challenging branches.

Note that although \tech{} is in theory language-independent, it still requires engineering efforts as the analysis and the parsing of test cases are highly specific to the chosen frameworks.
Thus we made the engineering effort to extend TELPA to support Java projects for the comparison.
Specifically, we adapted TELPA's program analysis components to work with Java code by incorporating JavaParser~\cite{javaparser}, allowing to effectively parse Java code. 
We also implemented the test execution process using the JUnit framework (the standard framework for unit testing in Java) and incorporated Jacoco~\cite{jacoco} for code coverage analysis (a widely-used tool for measuring Java code coverage). 
Additionally, we modified the processing of the LLM’s output to extract Java tests and address potential issues such as missing external libraries. 
In the future, we plan to evaluate \tech{} on the projects written in additional programming languages.
}

\delminor{
\textbf{Data Leakage.}
The projects used for evaluation may appear in the training data of the LLMs used in our study, which can cause the data leakage problem and bias the results~\cite{data_leak}. 
Particularly, following the guidelines provided by previous work~\cite{data_leak}, we also adopted three internal Java projects provided by our industrial partner (Huawei, a global leader company in IT), which can largely reduce the threat from the potential data leakage of open-source projects.
The three industrial projects have different functionalities, i.e., a program analysis toolkit, an online micro-service system, and a data analysis framework involving parallel computing and adaptation of design patterns. For ease of presentation, we refer to the three projects as PATool, Microservice and DAService in the following sections. Due to the company policy, we are unable to disclose further details.
We constructed \llmbase{}, which leverages \llmrevised{} for preceding test generation and
compared it with \llmrevised{}. We did not leverage \evo{} since the three industrial projects use Java 17, but \evo{} supports up to Java 11.
}

\delminor{
Table~\ref{tab:data_leakage} shows the comparison results for \llmbase{} vs. \llmrevised{} on each project in terms of branch coverage.
From the table, \llmbase{} achieves significantly higher branch coverage than \llmrevised{} across all three industrial projects. On average, \llmbase{} improves the branch coverage from 52.75\% to 69.71\% and the line coverage from 59.43\% to 72.71\%. 
The improvements of \tech{} on these unseen projects highlight its generalizability and effectiveness in real-world scenarios, independent of potential data leakage from open-source training data.
}

\textbf{Hallucination.}
Although \tech{} can significantly improve test coverage compared to the state-of-the-art SBST and LLM-based techniques, it is still unable to reach all the hard-to-cover branches within the given testing time budget.
Through our analysis, the major reason lies in the hallucination problem, namely LLMs confidently produce incorrect responses.
For example, some generated tests by \tech{} use non-existent parameters for invoking some methods.
Currently, several approaches have been proposed to relieve the hallucination problem in various software engineering tasks, such as fine-tuning LLMs with PPO~\cite{ppo} or DPO~\cite{dpo} algorithms based on high-quality data.
In the future, we can incorporate them to further improve the effectiveness of \tech{}.

\textbf{Orthogonality with Prompting Enhancement Methods.}
\tech{} is a novel prompting technique for addressing the challenge of hard-to-cover branches in software testing, which improves the prompt contents specific to this task.
Besides improving prompt contents, there are also some other prompting enhancement methods, including various CoT techniques (e.g.,  SCoT~\cite{scot} and CCoT~\cite{ccot}) and post-processing techniques (e.g., self-repairing~\cite{chatunitest}).
CoT techniques mainly elicit LLMs to produce intermediate reasoning steps based on the provided prompt for improving LLMs' effectiveness, while post-processing techniques mainly fix invalid code produced by LLMs based on the given prompt through static analysis or providing error messages to LLMs for self-repairing.
They are actually orthogonal to \tech{} (improving prompt contents), thus combining them may further improve test coverage.

\textbf{Assertion Generation.}
\tech{} focuses on improving branch coverage, a critical precursor to assertion generation.
Assertions generated may reveal bugs only if the buggy code is covered by tests.
While LLMs are capable of generating complete tests including both test prefixes and assertions~\cite{chattester, testpilot, our_empirical}, the quality of these assertions varies.
That is, some assertions can effectively determine whether the focal method behaves as expected with specific inputs, while some others are relatively simplistic (e.g., just checking if an object is \textit{None}).
Some works~\cite{toga, assertion_1, chen2016supporting} 
have specifically focused on assertion generation as part of test creation, and these techniques could potentially enhance the quality of LLM-generated assertions. In the future, we plan to refine these assertions by integrating assertion generation techniques and evaluating their effectiveness using metrics like bug detection and mutation scores~\cite{mutationscore}.

\ins{
\textbf{Trade-off in File-Level Analysis.}
\tool{} currently limits its analysis of method invocation sequences to the file hosting the method under test. This design represents a trade-off between comprehensiveness and computational efficiency.
The current file typically contains the most relevant usage scenarios for the target method, as it is both the definition location and often the location where it is the most frequently used~\cite{hits, chattester}. 
This local context provides sufficient information for generating effective tests in most cases. 
This approach of focusing on the local context is also widely used in many test generation approaches, such as CHATTESTER~\cite{chattester}, ChatUniTest~\cite{chatunitest}, and CODAMOSA~\cite{codamosa}.
While analyzing the entire codebase (e.g., looking at the other classes that might use the target class) could provide more comprehensive results, it would also introduce substantial computational costs. Additionally, cross-file analysis can be particularly challenging in dynamically typed languages like Python, where imports and method invocations are often resolved at runtime.
While limiting the analysis to the current file works well in many cases, we acknowledge that it may miss some usage scenarios, particularly when the target method is primarily used in other files. 
In the future, we can explore project-wide usage patterns to further improve our approach.
}

\del{
\textbf{Generalizability.}
From our study, the generalizability of \tech{} has been demonstrated to a certain extent by evaluating the performance on 27 projects and considering two types of preceding test generation tools (\sbst{} and \coda{}). 
Besides, the idea of \tech{} is not specific to Python projects (our evaluation benchmark). The involved program analyses in \tech{} can be extended to the projects developed in other programming languages, as long as similar data and control flow analyses of this study can be performed on those languages. 
Particularly, \tech{} has been deployed for Java projects in Huawei (our industry partner).
In fact, \tech{} is not restricted to using a certain LLM, and many code generation-related LLMs~\mbox{\cite{codellama,starcoder, codegeex}} support program understanding across various programming languages.
In the future, we will evaluate \tech{} with diverse LLMs on the projects in other programming languages.
}

\section{Related Work}
\label{sec:related_work}

We present traditional and deep-learning-based test generation techniques as related work.
The most widely-studied traditional test generation technique category is SBST, including Pynguin~\cite{pynguin} investigated in our work. Pynguin is a state-of-the-art technique for Python projects, and there are also SBST techniques for Java projects, such as Randoop~\cite{randoop} and EvoSuite~\cite{evosuite}. Randoop performs random generation, while EvoSuite leverages more advanced search algorithms to guide test generation. They have also been improved by incorporating string literals from human-written tests~\cite{testminer} and object construction graphs~\cite{evoobj}.

Some other techniques, instead, employ symbolic execution to improve test coverage~\cite{pex, symbolic0,combine_symbolic1,combine_symbolic2, bidirection_symbolic}.
For example, 
Galeotti et al.~\cite{symbolic0} integrated dynamic symbolic execution (DSE) into the Genetic Algorithm (GA) adopted by EvoSuite. Their approach mutates tests, where primitive values can influence the fitness, to cover specific corner cases without sacrificing general coverage.
Braione et al.~\cite{combine_symbolic1} and Li et al.~\cite{combine_symbolic2} transformed path conditions into optimization problems and solved them with SBST and machine learning techniques, respectively.
Baluda et al.~\cite{bidirection_symbolic} combined symbolic execution and symbolic reachability analysis to improve the effectiveness by testing rare execution conditions and eliminating infeasible branches.
However, it is well-known that symbolic execution has difficulties in dealing with complex data types and objects, thus unable to address our specific challenges. %

Our approach \tech{} utilizes program-analysis-enhanced prompting to exploit the code comprehension ability of LLMs and facilitate the effective test generation for hard-to-cover branches. The above-mentioned techniques, while not effective for this task, can still be incorporated into \tech{} as preceding testing tools and benefit from our new prompting method.

There are also several deep-learning-based test generation techniques~\cite{athenatest,a3test}.
Before prompting LLMs for test generation, 
ATHENATEST~\cite{athenatest} built a transformer model based on a large dataset of target methods and tests for test generation. 
A3Test~\cite{a3test} first built a pre-trained language model for assertions in a self-supervised manner based on PLBART~\cite{plbart}, and then fine-tuned it with test generation data. 
CAT-LM~\cite{cat_lm} trained a GPT-style LLM on Java and Python repositories, using a unique objective that maps source code to corresponding test files.
Recently, some work leveraged LLMs for test generation~\cite{codamosa,chatunitest,chattester, test_gen, testpilot, teut}, including \coda{} and \tester{} used in our study.
For example, 
TestGen-LLM~\cite{test_gen} utilized assured offline LLM-based software engineering~\cite{assuredllm} to integrate language models as a service within a comprehensive software engineering workflow, ultimately recommending unit tests with higher coverage. 
ChatUniTest~\cite{chatunitest} leveraged LLMs to generate tests with similar prompt contents used in \coda{}, but designed post-processing strategies to fix invalid tests generated by LLMs.

\tech{} is also a LLM-based test generation technique, which improves prompt contents with the aid of program analysis (i.e., contextual information purification specific to the challenges of hard-to-cover branches).
On one hand, those test-repair strategies can be combined with \tech{} to further improve its performance.
On the other hand, these deep-learning-based techniques can be also used as preceding test generation tools in \tech{} for orthogonal integration.

There are also some empirical evaluations focusing on LLM-based test generation.
Wang et al.~\cite{empirical1} surveyed 102 recent papers on using LLMs for software testing, providing a comprehensive overview.
Schäfer et al.~\cite{testpilot} conducted an extensive study evaluating the effectiveness of using API signatures, documentation, and related information for prompting LLMs to generate unit tests.
Tang et al.~\cite{empirical2} presented a systematic comparison of test suites generated by ChatGPT and EvoSuite.
Yang et al.~\cite{our_empirical} conducted the first empirical study to investigate the unit test generation effectiveness of open-source LLMs.
Different from these empirical studies, \tech{} is a novel LLM-based test generation technique with the aid of program analysis.

\section{Conclusion}

Existing test generation techniques face difficulties in covering branches involving specific complex objects and intricate inter-procedural dependencies. %
To address this issue, we propose a novel technique, named \tech{}. \tech{} leverages program analysis %
to assist LLMs in constructing complex objects and understanding code semantics, thus improving the test generation performance. 
Additionally, \tech{} employs counter-example sampling and coverage-based feedback to guide LLMs to  effectively and efficiently  generate tests. 
Our experiments on 27 open-source Python projects \insminor{, four open-source Java Projects, and three internal Java projects} demonstrate that \tech{} significantly outperforms both the state-of-the-art SBST and LLM-based techniques.
We released our implementation of \tech{} and experimental data for replication and practical use. Please find them at \delminor{\textbf{\url{https://anonymous.4open.science/r/TELPA-CF78}}}\insminor{\textbf{\url{https://zenodo.org/records/15410112}}}.

\begin{acks}
\insminor{
We thank the editor and the anonymous reviewers for their constructive suggestions to help improve the quality of this paper. 
This work was supported by the National Natural Science Foundation of China (Grant Nos. 62322208, 62232001), and Huawei Fund.
}
\end{acks}

\normalem
\bibliographystyle{ACM-Reference-Format}
\bibliography{9_ref}

\end{document}